\documentclass{article}

\usepackage{arxiv}

\usepackage[utf8]{inputenc} 
\usepackage[T1]{fontenc}    
\usepackage{hyperref}       
\usepackage{url}            
\usepackage{booktabs}       
\usepackage{amsfonts}       
\usepackage{nicefrac}       
\usepackage{microtype}      
\usepackage{lipsum}		
\usepackage{graphicx}
\usepackage{natbib}
\usepackage{doi}
\usepackage{caption}
\usepackage{subcaption}
\usepackage{algorithm}
\usepackage{algorithmic}

\title{On The Power of Subtle Expressive Cues in the Perception of Human Affects}

\author{  Ezgi Dede\\
College of Engineering\\ 
	Koc University, Istanbul, Turkey \\
	\texttt{edede19@ku.edu.tr} \\
 	\And
  Kamile Asli Agilonu\\
College of Engineering\\ 
	Koc University, Istanbul, Turkey \\
	\texttt{kagilonu14@ku.edu.tr} \\
 	\And
\href{https://orcid.org/0000-0003-3618-4166}{\includegraphics[scale=0.06]{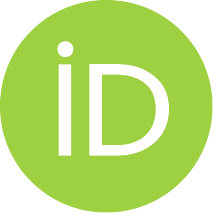}\hspace{1mm}Ergun Akleman}\thanks{Joint with Computer Science and Engineering Department.} \\
	Visual Computing \& Computational Media, School of PVFA\\
 Computer Science and Engineering, College of Engineering\\
 Texas A\&M University, College Station, TX, 77831\\
	\texttt{ergun@tamu.edu} \\
 	\And
	 \href{https://orcid.org/0000-0002-1524-1646}{\includegraphics[scale=0.06]{orcid.pdf}\hspace{1mm}Metin Sezgin}\\
	 College of Engineering\\ 
	Koc University, Istanbul, Turkey \\
	\texttt{mtsezgin@ku.edu.tr} \\
}

\renewcommand{\shorttitle}{Power of Minor Details}

\hypersetup{
pdftitle={A template for the arxiv style},
pdfsubject={q-bio.NC, q-bio.QM},
pdfauthor={David S.~Hippocampus, Elias D.~Striatum},
pdfkeywords={First keyword, Second keyword, More},
}

\begin{document}
\maketitle

\begin{abstract}
In this study, we introduce a sketch-based method for testing how subtle expressive cues influence the perception of affect in illustrations of human figures. We specifically study the impact of human posture and gaze direction, implicitly specified through nose orientation, on perceived emotions and mood. Through a series of user studies using sketchy illustrations of a running figure, where a professional illustrator manipulated gaze direction through adjustments on the nose orientation, we found that this simple change resulted in a diverse range of perceived affects, spanning from fear to concern and wonder. These findings shed light on the importance of fine details in defining context for context-aware system designs and underscore the importance of recognizing and expressing affect. Understanding minor expressive cues is crucial to developing emotionally intelligent systems capable of expressing affect.
\end{abstract}

\section{Introduction and Motivation}

Understanding how to perceive human affects such as feeling, emotion, attachment, or mood from affect displays is critical for a wide range of applications, from affective computing and context-aware systems to visual storytelling. 
\begin{figure*}[htb] 
\centering
\includegraphics[width=0.19\linewidth]{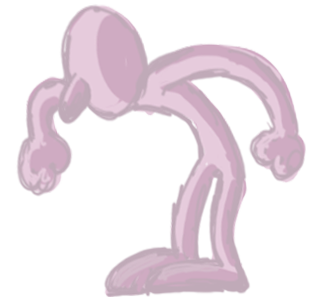}
\includegraphics[width=0.19\linewidth]{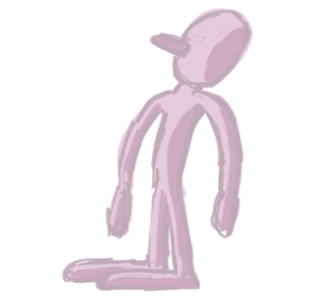}
\includegraphics[width=0.19\linewidth]{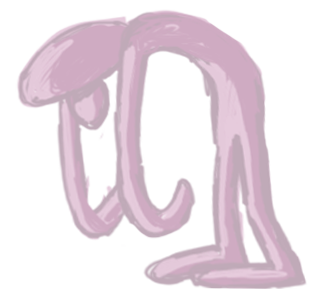}
\includegraphics[width=0.19\linewidth]{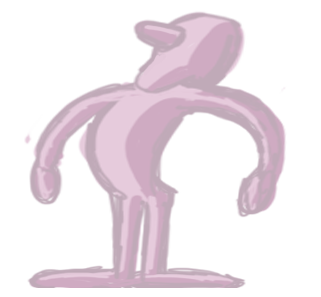}
\includegraphics[width=0.19\linewidth]{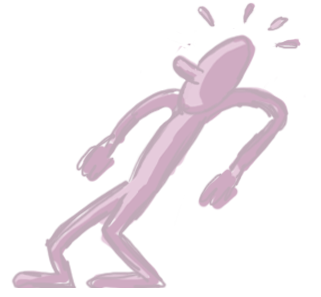}
\caption{Expressions such as anger, happiness, sadness, pride, and fear only with posture without any facial expression.}
\label{images/posture}
\end{figure*}
Affective displays can be any type of expressive cues, such as facial, vocal, or gestural behavior, that serves as an indicator of affect \cite{vandenbos2007apa}. 
Unfortunately, in the scientific literature, only facial, vocal, and gestural expressions are studied \cite{ekman1979facial, ekman1997universal, ekman1999facial, russell2003facial}. However, expressions can go beyond facial, vocal, and gestural expressions. As shown in Figure~\ref{images/posture} even body posture can be sufficient to convey human affects \cite{liu2012, akleman2015}. Using standard scientific processes, it is hard to classify and categorize such unusual and subtle expressions. In this work, we have developed a methodology to evaluate such unusual and subtle expressions. Using this methodology, we have demonstrated that the perception of human affects can be significantly changed even by subtle expressive cues of affective displays, such as gaze direction. 

\subsection{Context \& Motivation} 

\begin{figure}[hbtp]
\centering
\includegraphics[width=0.99\linewidth]{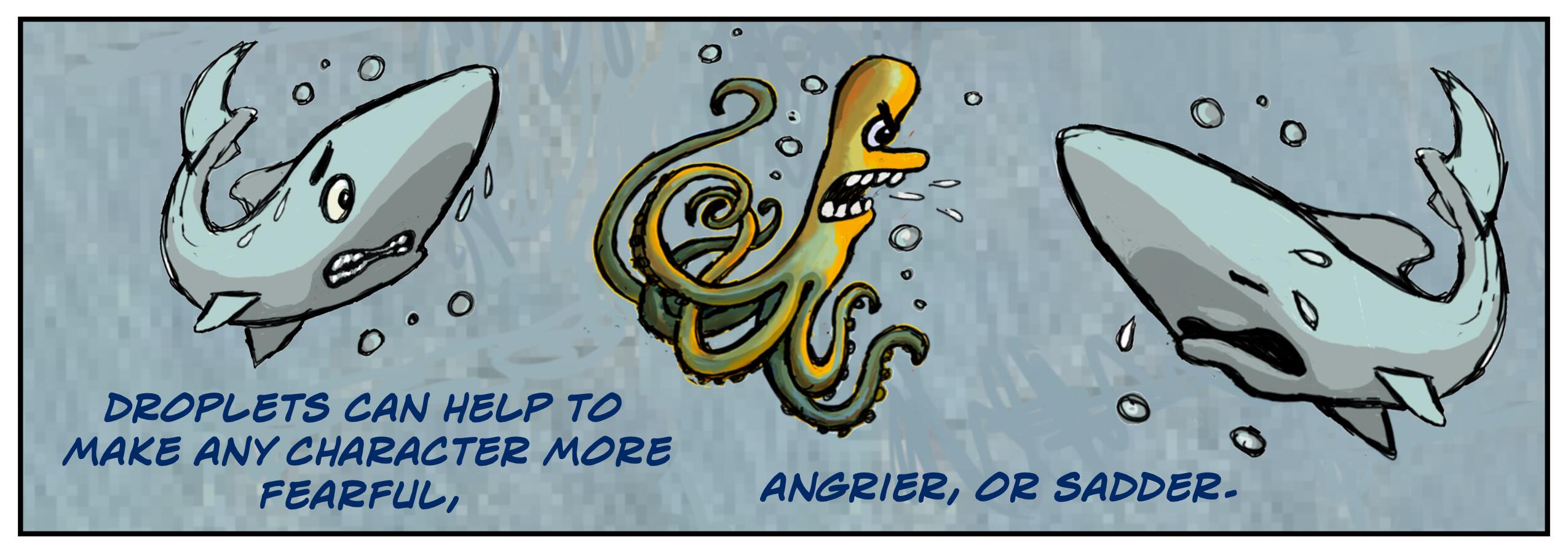}
    \caption{Droplets help to change and exaggerate perceived expressions \cite{akleman2020}.  }
\label{droplets/0}
\end{figure}

The main problem with understanding human affects is that expressive cues of affective displays can be subtle and context-dependent. These cues can come from a wide range of sensory input. To understand human affects, it is important to consider all types of sensory input, even if a cue appears to be relatively minor. It is also important to process these cues in their context to recognize the underlying human affect. Small details such as the addition of a few droplets have already been shown to change and exaggerate the expression of human affects \cite{akleman2020} (see Figure~\ref{droplets/0}). Therefore, every cue from a small detail to overall body posture can be crucial in understanding human affects by providing a rich contextual expression.

Unfortunately, the interpretation of subtle cues is not yet well understood in the scientific literature.  Expert actors, comic book artists, cartoonists, and animators know how to create subtle and context-aware expressive cues effectively. However, the knowledge of these artists is gained through apprenticeship training obtained by interacting with other experts. Only a handful of artists formally describe their methods \cite{johnston1981illusion, mccloud1993understanding, blair1995cartoon, mccloud2006making, eisner2008comics, eisner2008graphic, celik2011on}. User study procedures that are common in the scientific community are too complex to discover the impact of a wide variety of subtle context-aware cues. Simple procedures are needed to discover the effects of subtle cues. For this goal, we have developed a middle ground by combining expert knowledge with very simple user studies to quickly test information provided by experts. We develop user studies with extremely simple sketch-based illustrations. This simplifies the creation of the data set. On the other hand, we work with expert cartoonists to create the data set. Therefore, we can be sure that the set is in good condition for user studies.

\begin{figure}[hbtp]
\centering
\includegraphics[width=0.99\linewidth]{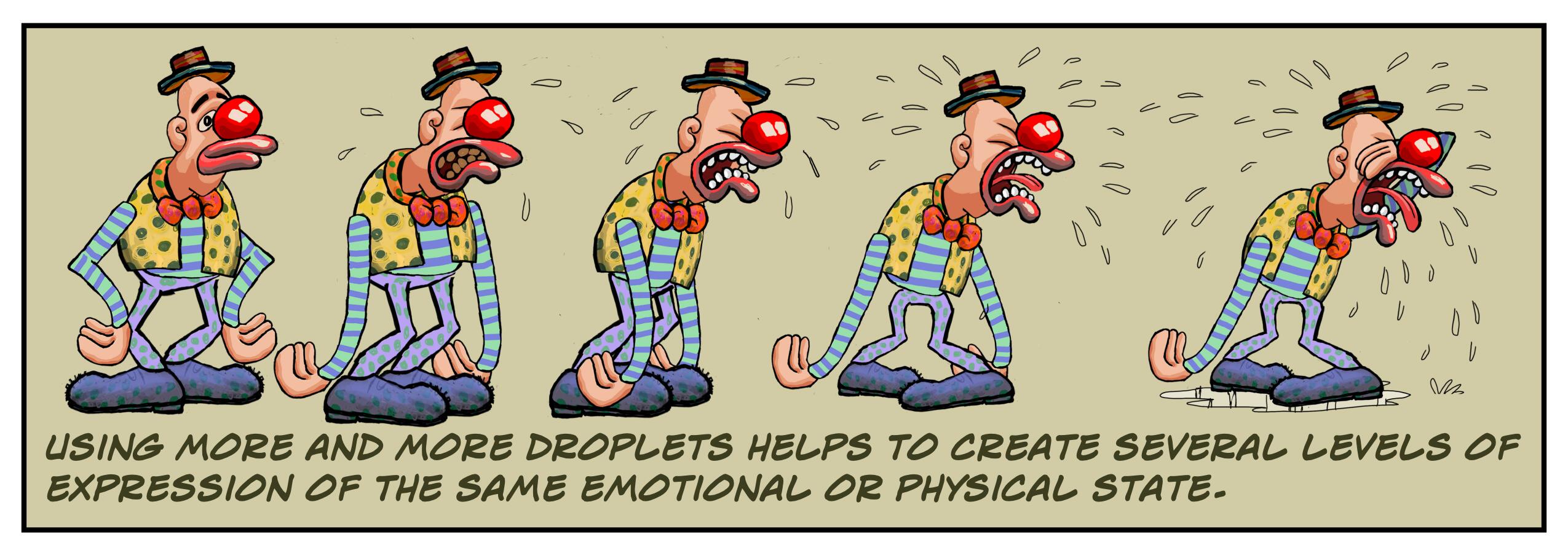}
    \caption{This example shows that props such as droplets can be used to significantly increase the intensity of perceived expressions. Droplets near the eyes create a sad expression, droplets on the forehead give the impression of fatigue, and droplets under the nose give the impression of sickness even in a neutral face. For more examples see \cite{akleman2020}.  }
\label{droplets/1}
\end{figure}

\subsection{Basis and Rationale}\label{sec:basis}
\label{BasisAndRationale}

Understanding subtle expressions of human affects in their context is crucial in human interactions. 
Individuals are expected to behave within certain norms required by the context, such as gatherings or work environments. As the context also depends on the culture, we can see problematic interactions between people originating from different cultures. A person coming from a culture that has strict boundaries of personal space would feel uncomfortable in an environment where people do not acknowledge such boundaries. Thus, it is important for any computational agent working with humans to analyze and understand the context in which they are working. This raises the question: What is the context? In ubiquitous computing, a formal description is given as "physical activity, location, and the psychophysiology and effective state of a person" \cite{bulling2011}.

In the field of human-computer interaction, context-aware systems are frequently used because they offer easy use to users when they adjust their functionalities to the given context. An example would be an intelligent system designed to receive voice commands during emergencies. For example, in the event of a house fire, it is expected that the caller will speak louder and faster than usual. If the system is not designed to adapt to this situation and understand the user, the consequences of this failure would be critical.

\subsection{Contributions}

In this work, we develop a formal evaluation methodology to evaluate the impact of subtle expressive cues. Our methodology is based on methodical sketches drawn by expert illustrators. Using these sketches, we study the effect of a particular expressive cue in detail. Using user studies, we can validate or reject a hypothesis on the effect of a particular subtle expressive cue. 

In this work, we also provide a proof of concept for the method. using our methodology, we have demonstrated that a relatively subtle change in a drawing can make a significant change in the context created by the person's effective state. As a result, we argue that context-aware systems should be designed to detect and interpret subtle expressive cues, since small elements can create a completely different context. 

\begin{figure}[hbtp]
\centering
     \begin{subfigure}[t]{0.49\linewidth}
\includegraphics[width=0.49\linewidth]{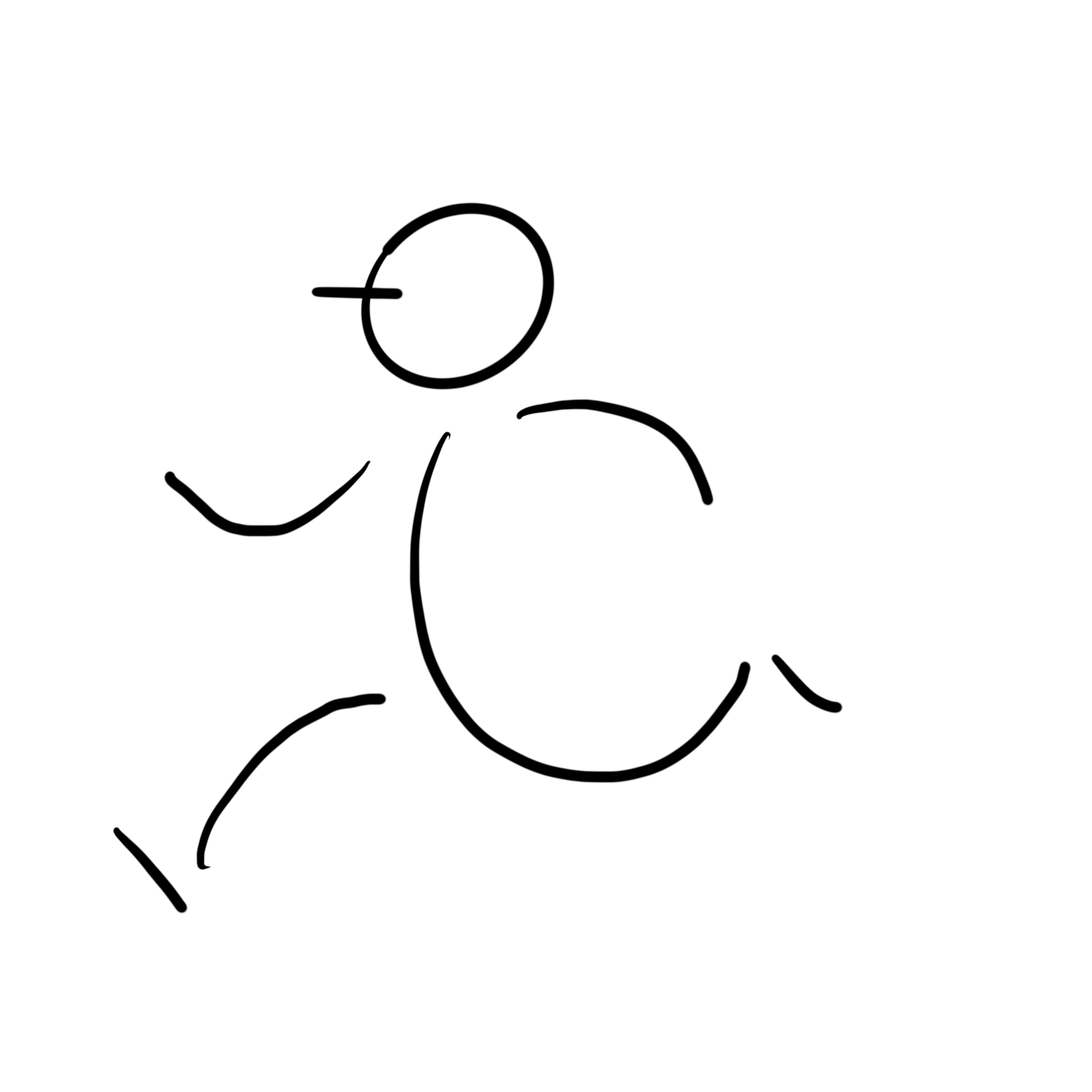}
\includegraphics[width=0.49\linewidth]{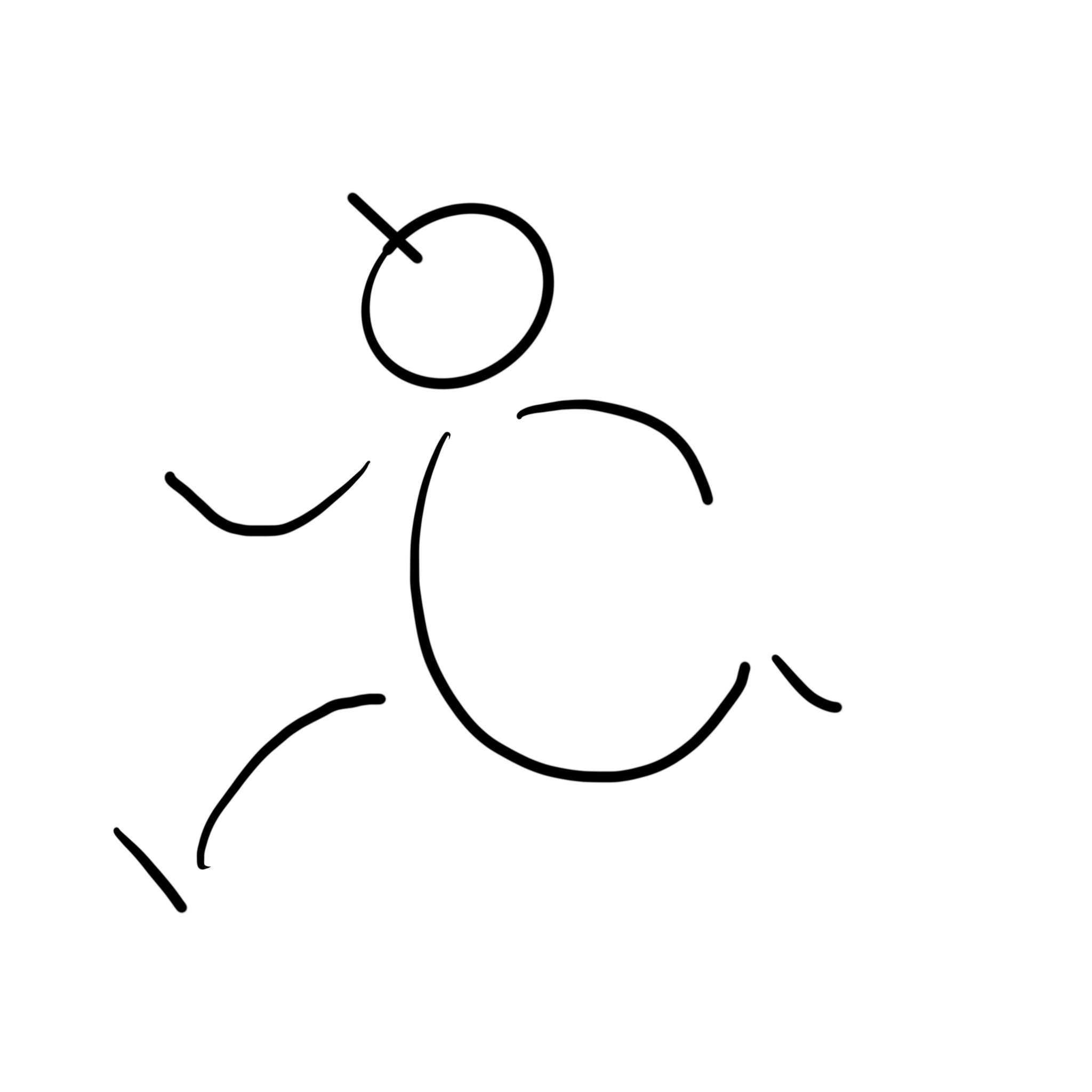}
    \caption{\footnotesize Set-0: A wire person (see Figure \ref{images/0} for the full set). }
    \label{fig_set0}
 \end{subfigure}
 \hfill
     \begin{subfigure}[t]{0.49\linewidth}
\includegraphics[width=0.49\linewidth]{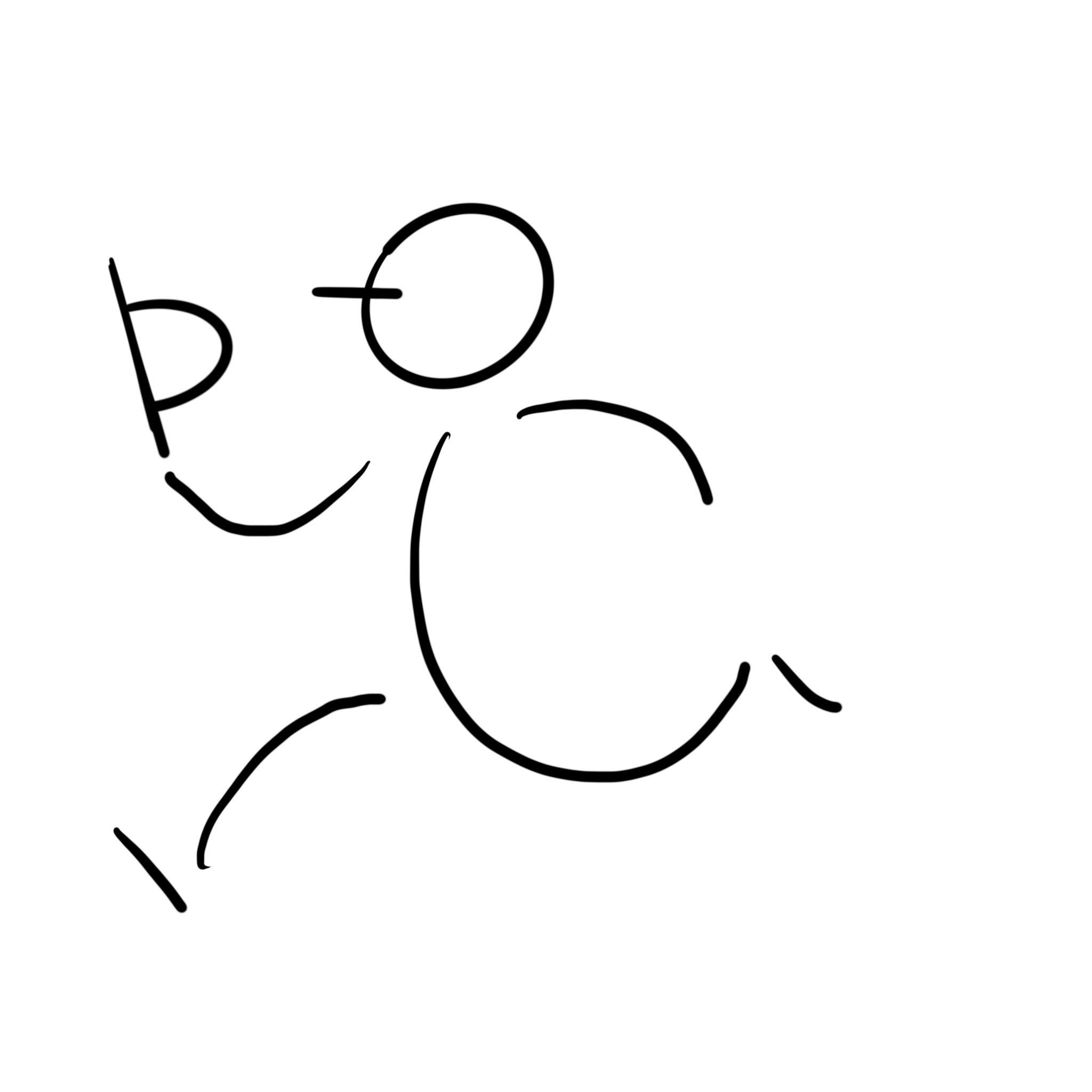}
\includegraphics[width=0.49\linewidth]{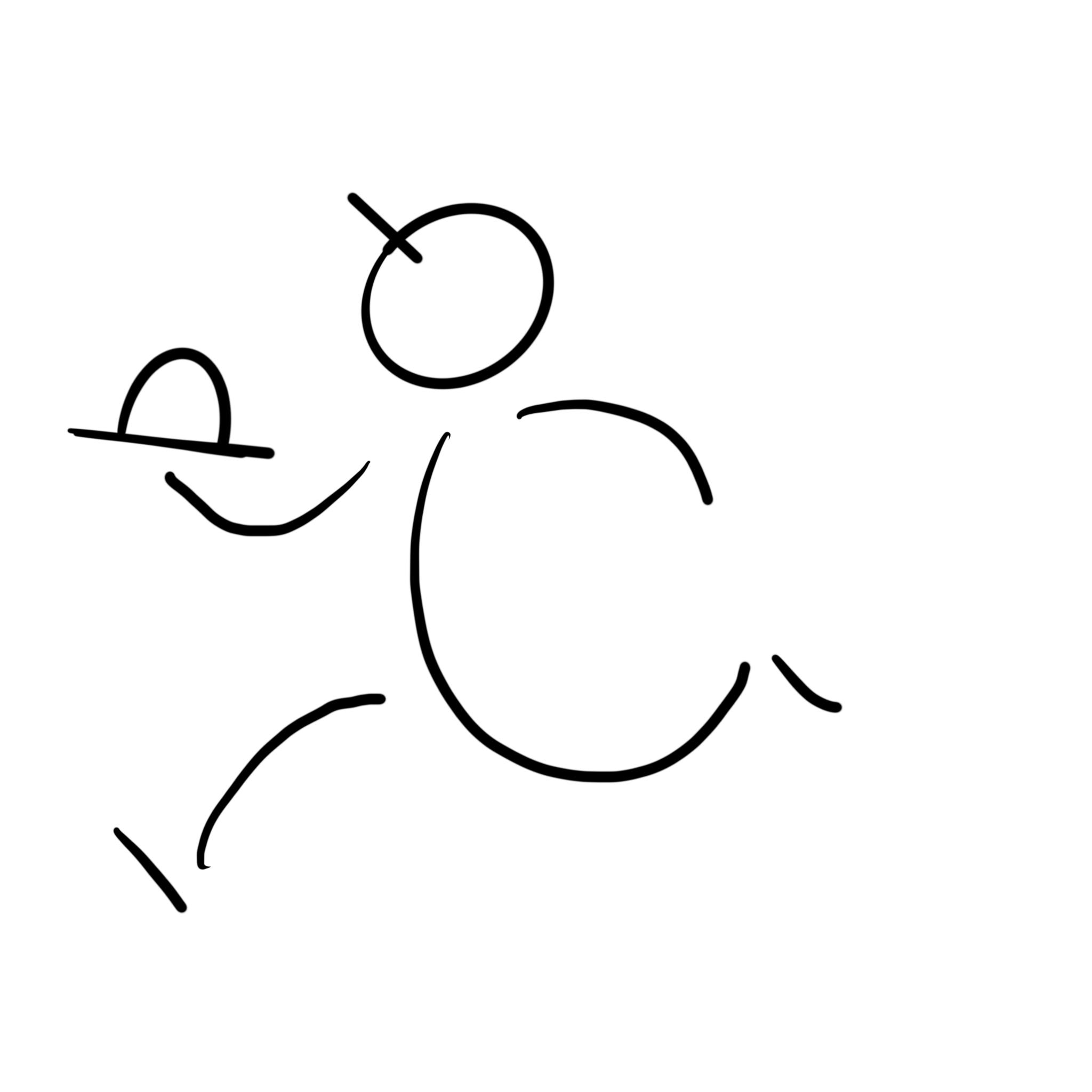}
    \caption{\footnotesize Set-1: A wire person with a hat (see Figure \ref{images/1} for a complete set). }
    \label{fig_set1}
 \end{subfigure}
 \hfill
     \begin{subfigure}[t]{0.49\linewidth}
\includegraphics[width=0.49\linewidth]{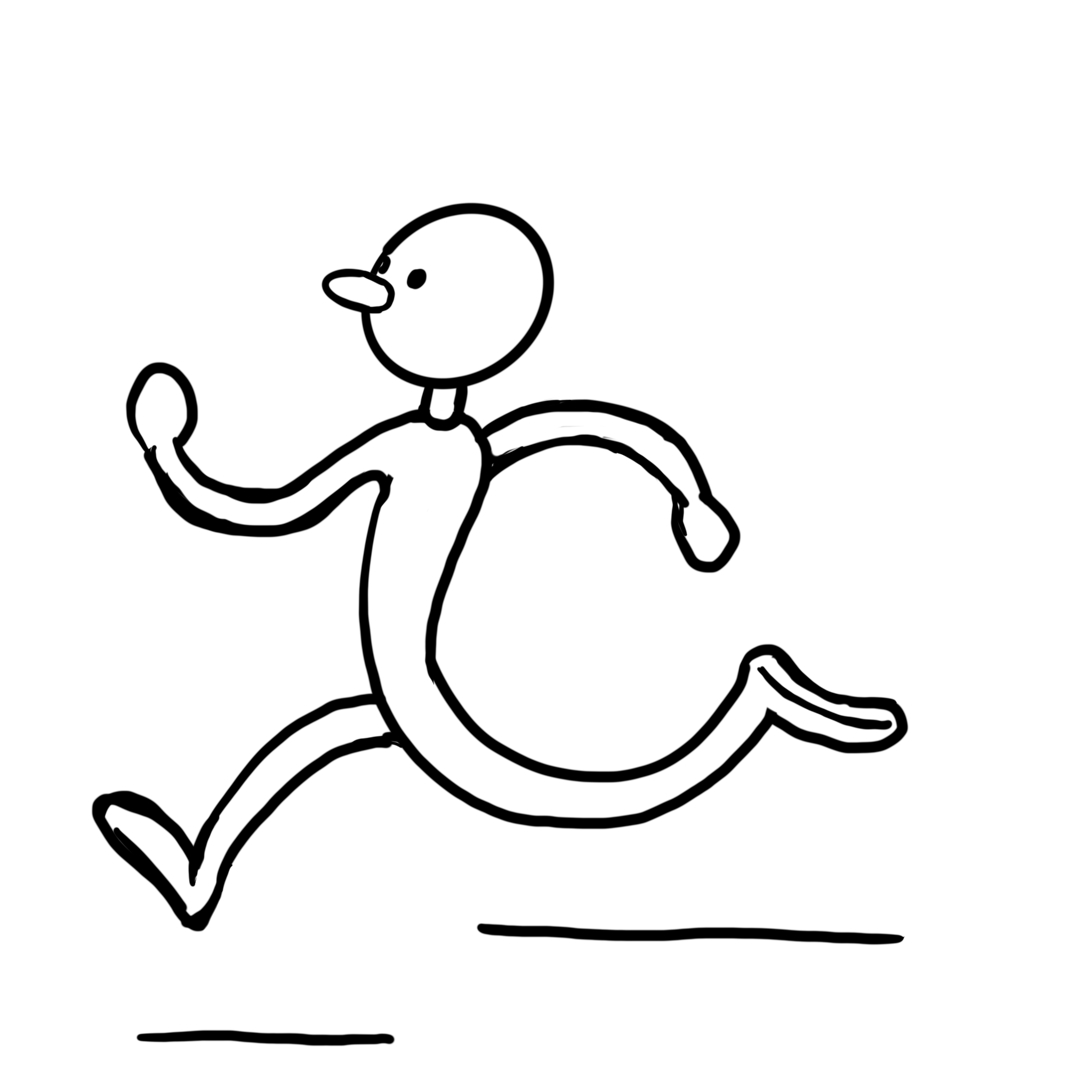}
\includegraphics[width=0.49\linewidth]{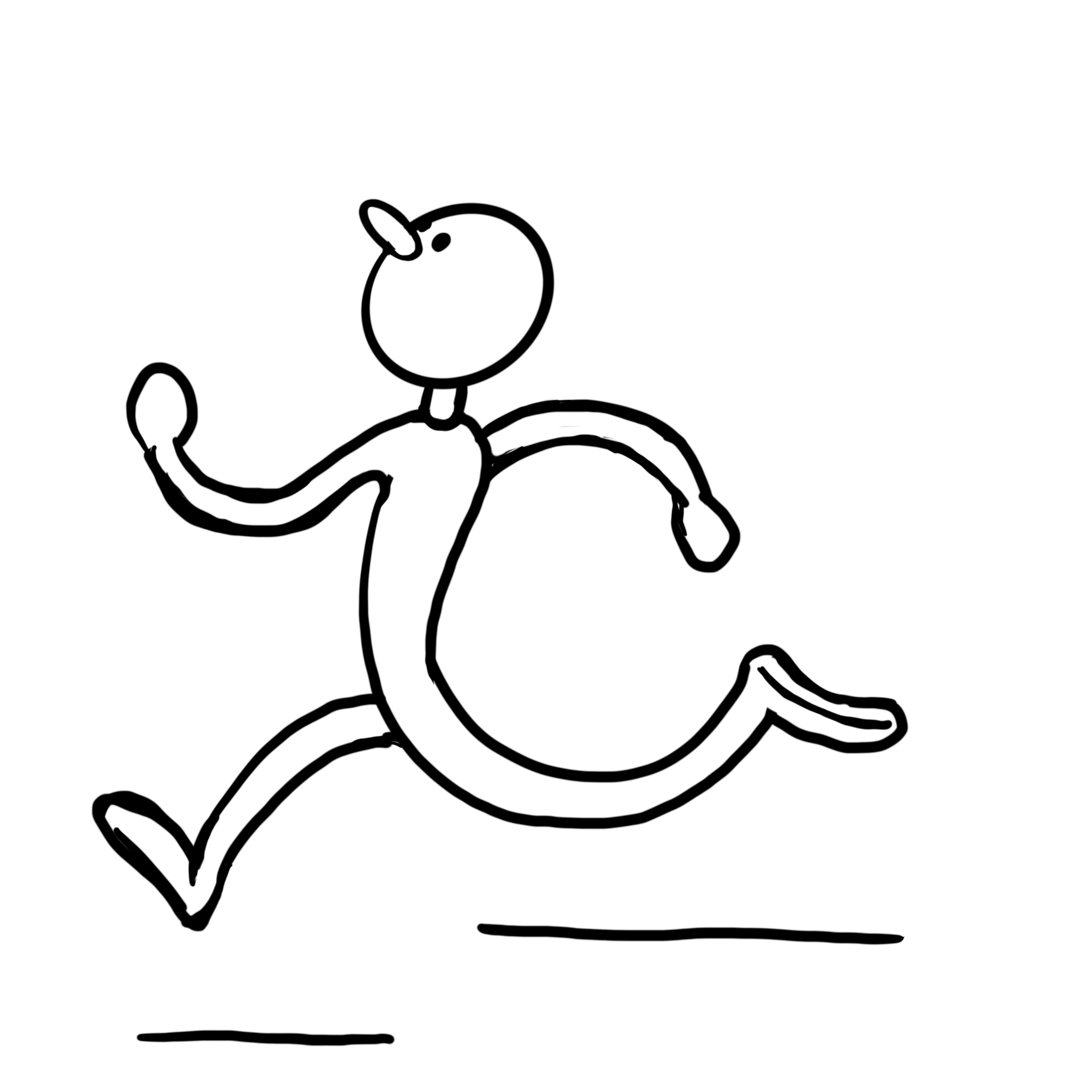}
    \caption{\footnotesize Set-2: A solid person slightly behind viewpoint (see Figures \ref{images/4a} and \ref{images/4b} for the full set). }
    \label{fig_set2}
 \end{subfigure}
 \hfill
     \begin{subfigure}[t]{0.49\linewidth}
\includegraphics[width=0.49\linewidth]{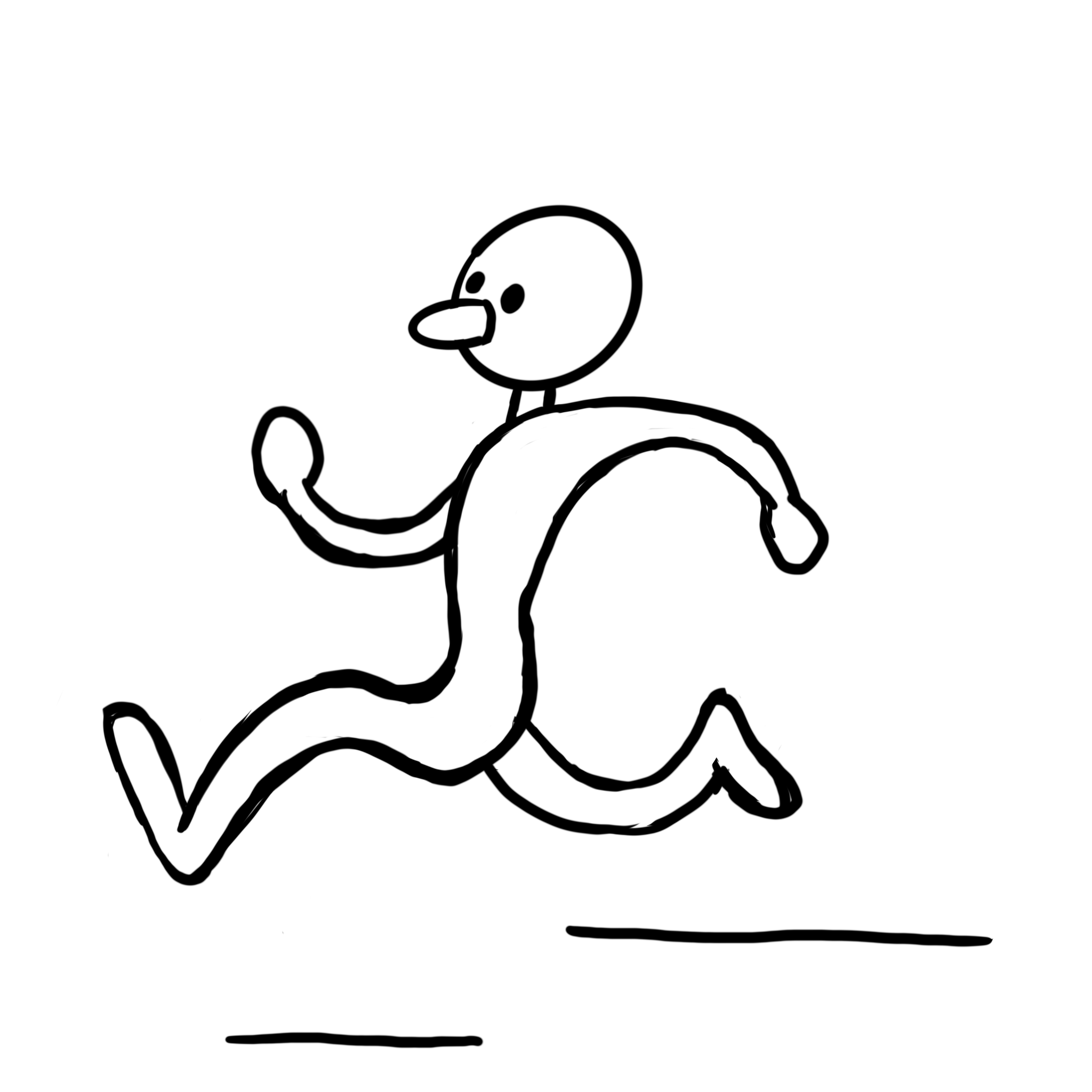}
\includegraphics[width=0.49\linewidth]{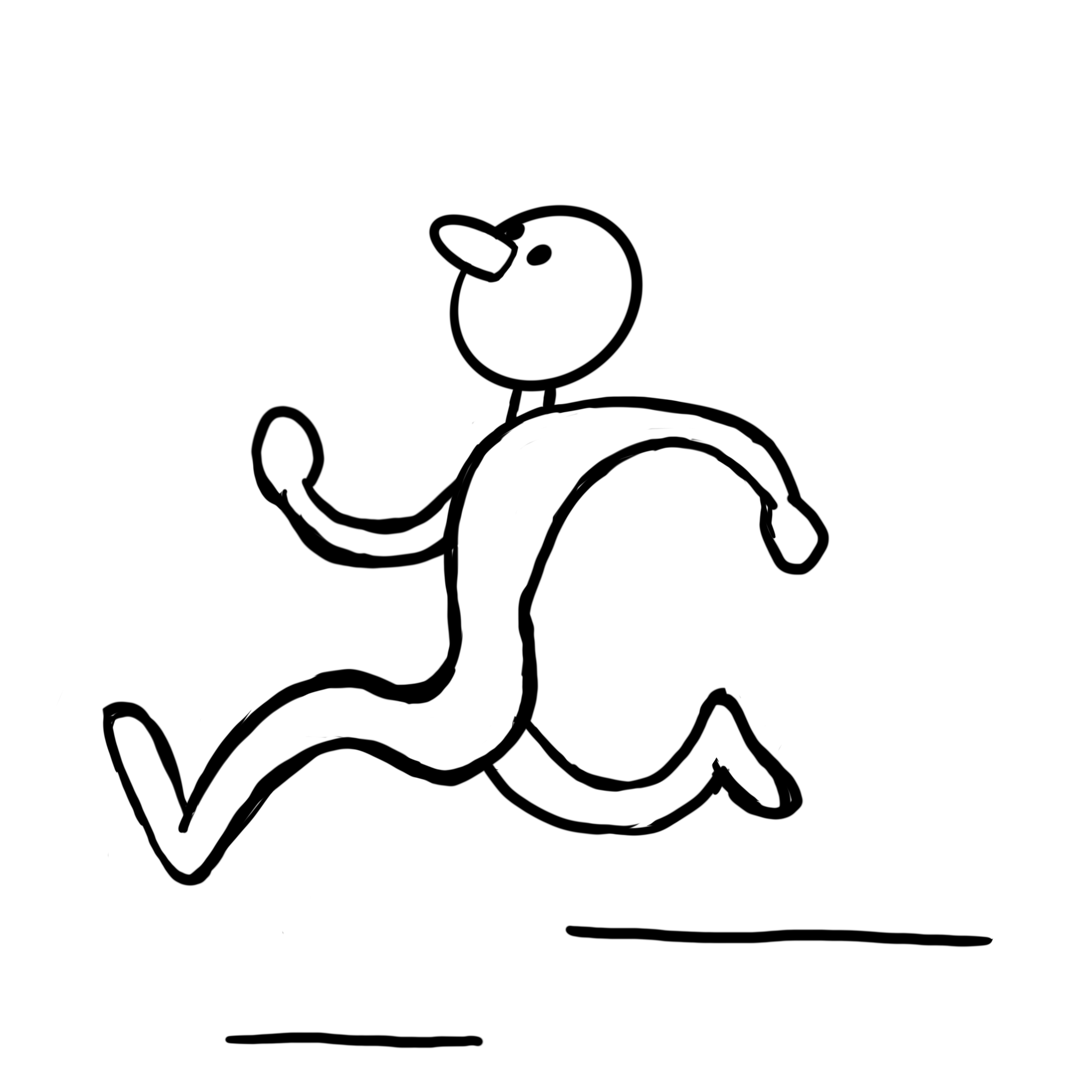}
    \caption{\footnotesize Set-3: A solid person with side view (see Figure \ref{images/3} for the complete set). }
    \label{fig_set3}
 \end{subfigure}
 \hfill
    \caption{An artist created sets of sketchy illustrations to test the validity of our hypothesis. These are some examples of four sets that provide simplified visual depictions of running people.  }
\label{images/examples}
\end{figure}

To run the user study, an expert illustrator initially created four different sets of illustrations with different levels of visual detail (see Figure~\ref{images/examples} for two samples from each set). 
For this study, by discussing with the illustrator, we chose to use the solid drawings shown in the figures of the sets~\ref{fig_set2} and ~\ref{fig_set3} because they provided less ambiguity and their body components were proportionally more accurate and robust. They also had eyes, which reduced the ambiguity of the head of the figure. Finally, we chose to use the side view shown in Figure~\ref{fig_set2} and added the gaze direction to further reduce the ambiguity. 

\begin{figure*}[htb] 
\centering
\begin{subfigure}[t]{0.99\linewidth}
\includegraphics[width=0.32\linewidth]{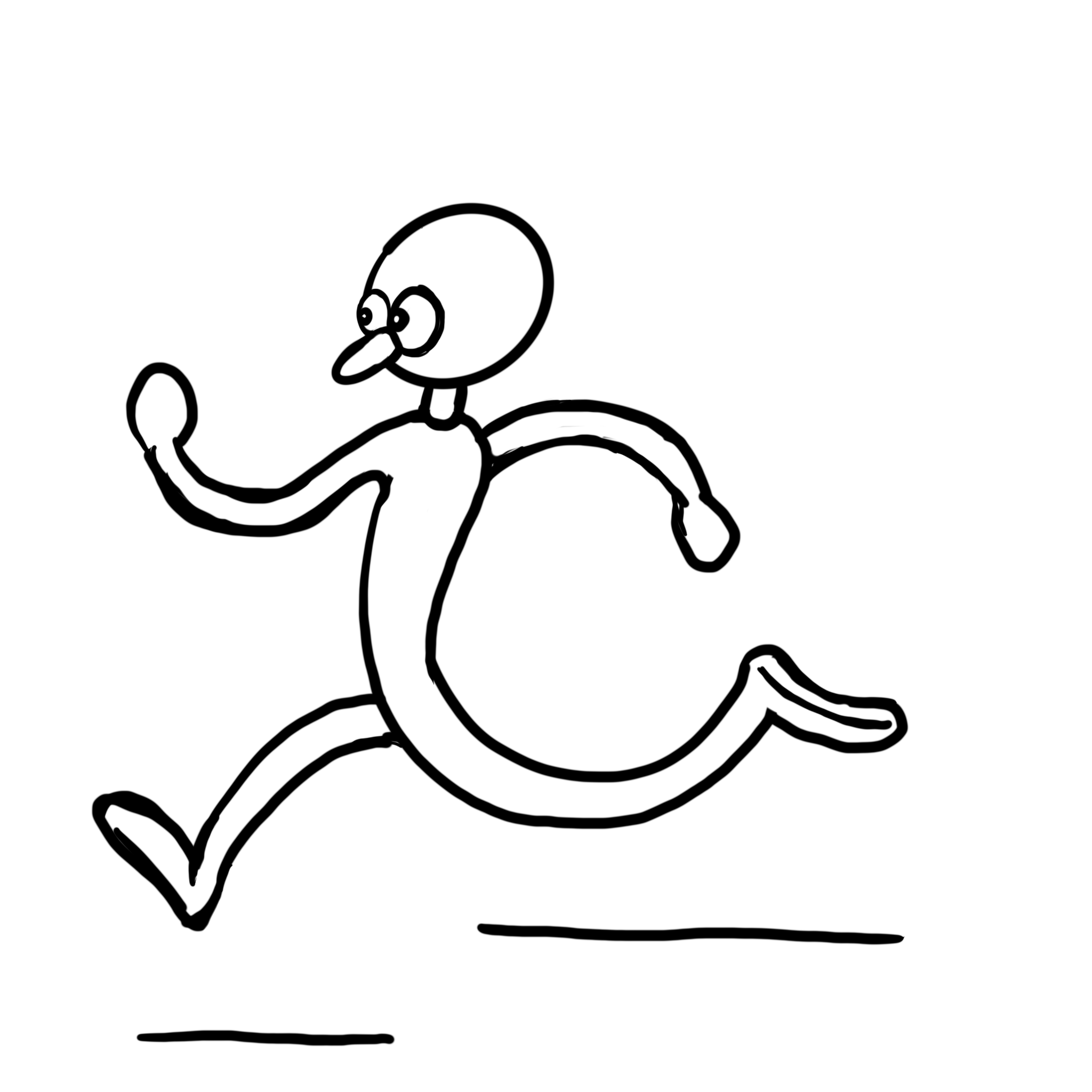}
\includegraphics[width=0.32\linewidth]{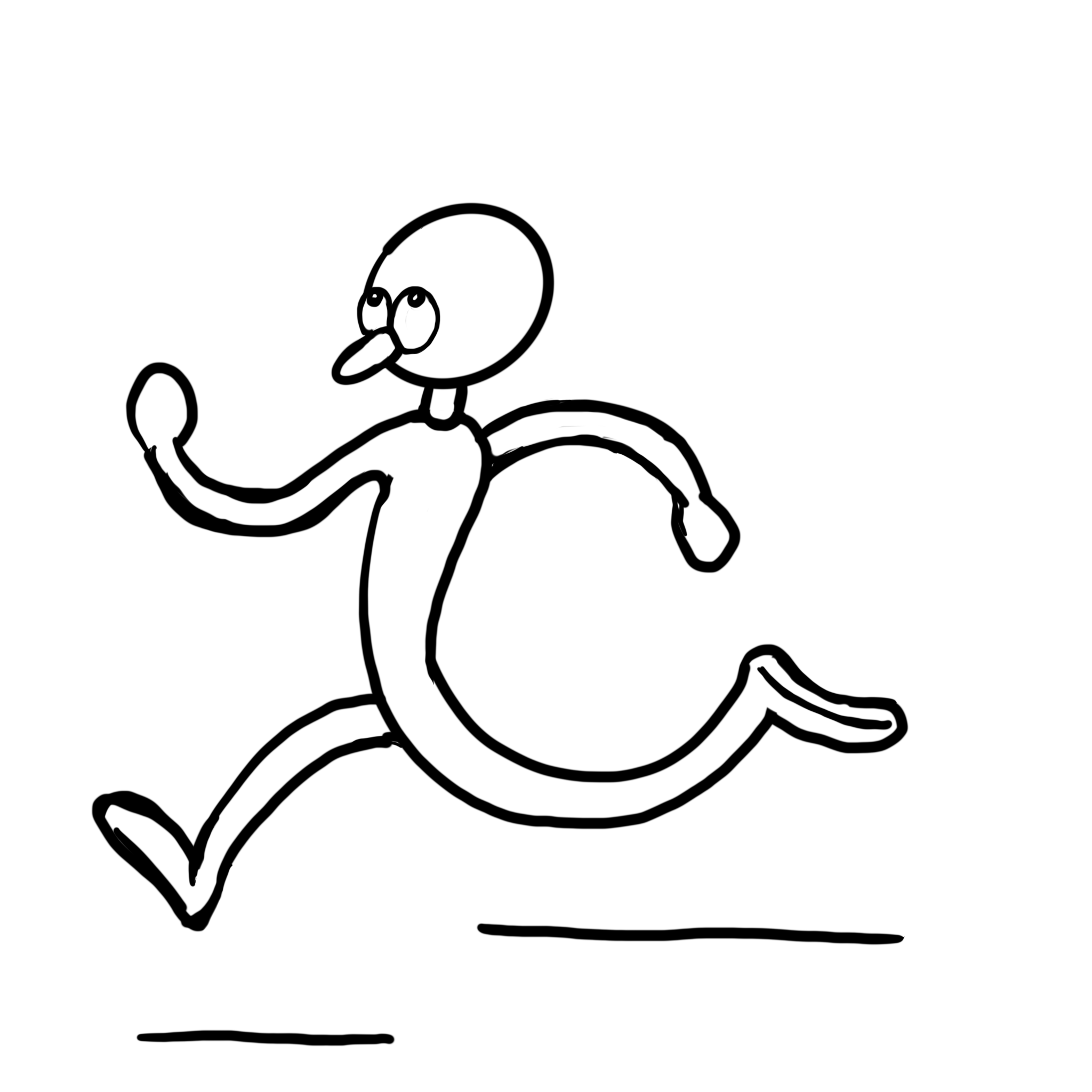}
\includegraphics[width=0.32\linewidth]{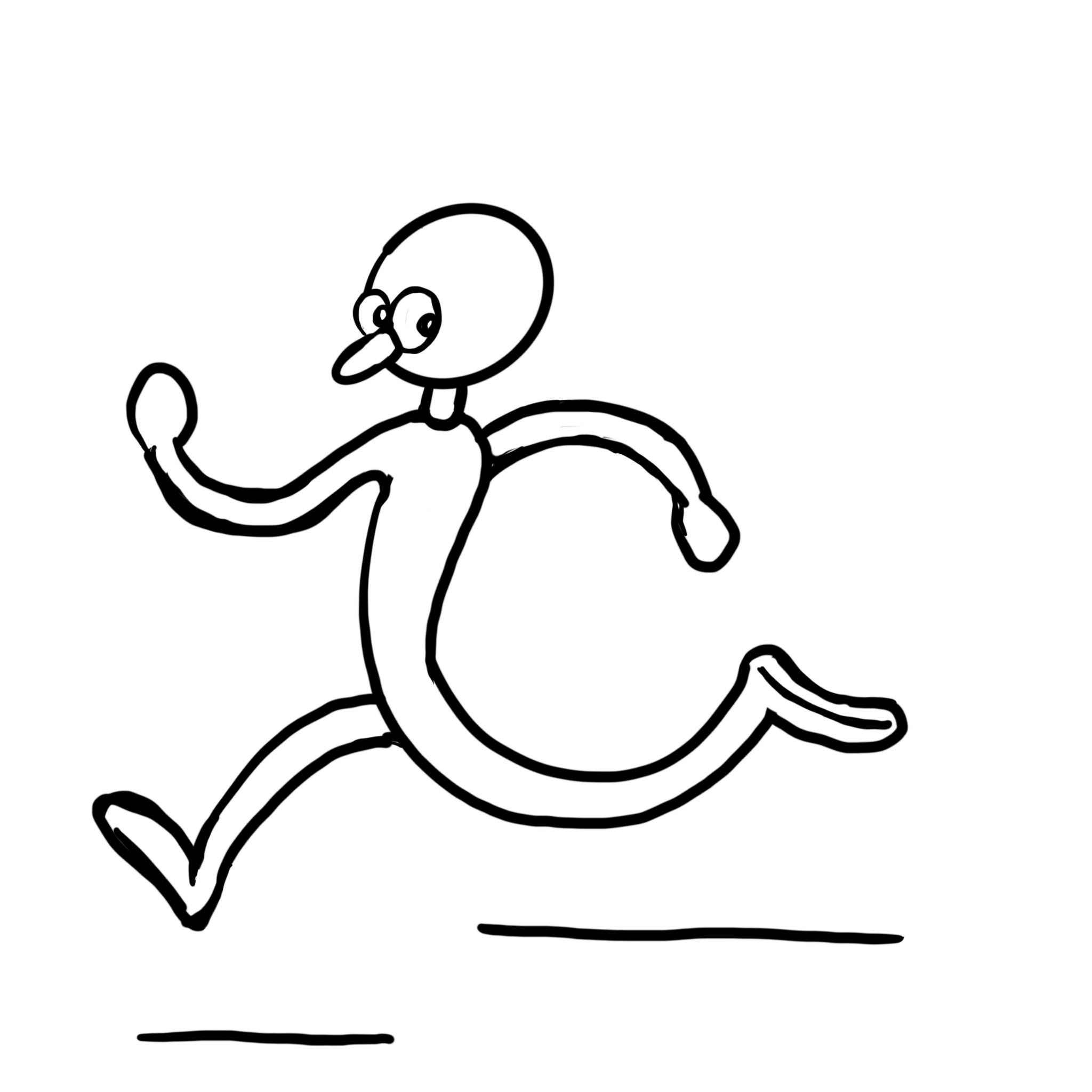}
\caption{\footnotesize Facing down and front. }
\label{fig_4/0}
 \end{subfigure}
 \hfill
\begin{subfigure}[t]{0.99\linewidth}
\includegraphics[width=0.32\linewidth]{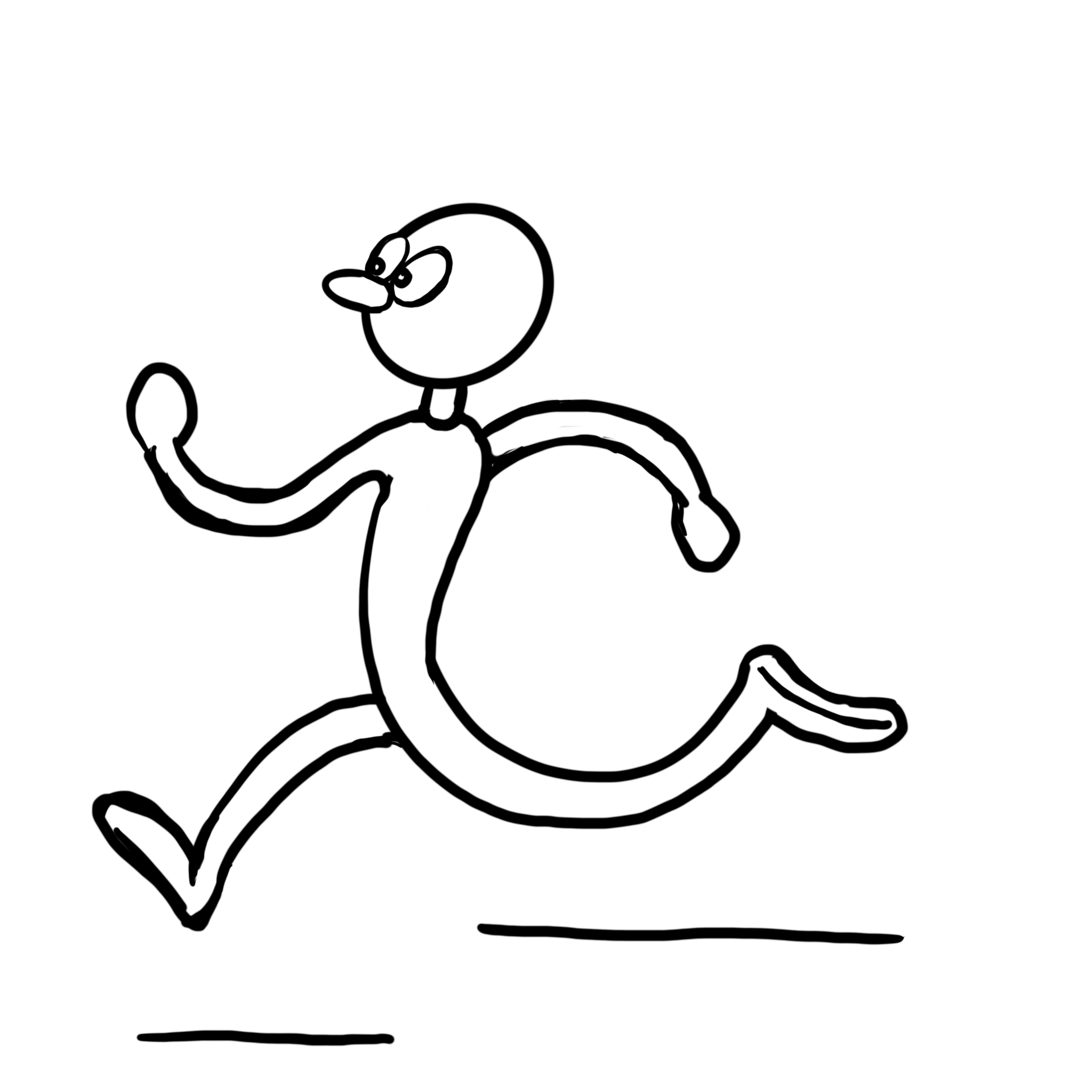}
\includegraphics[width=0.32\linewidth]{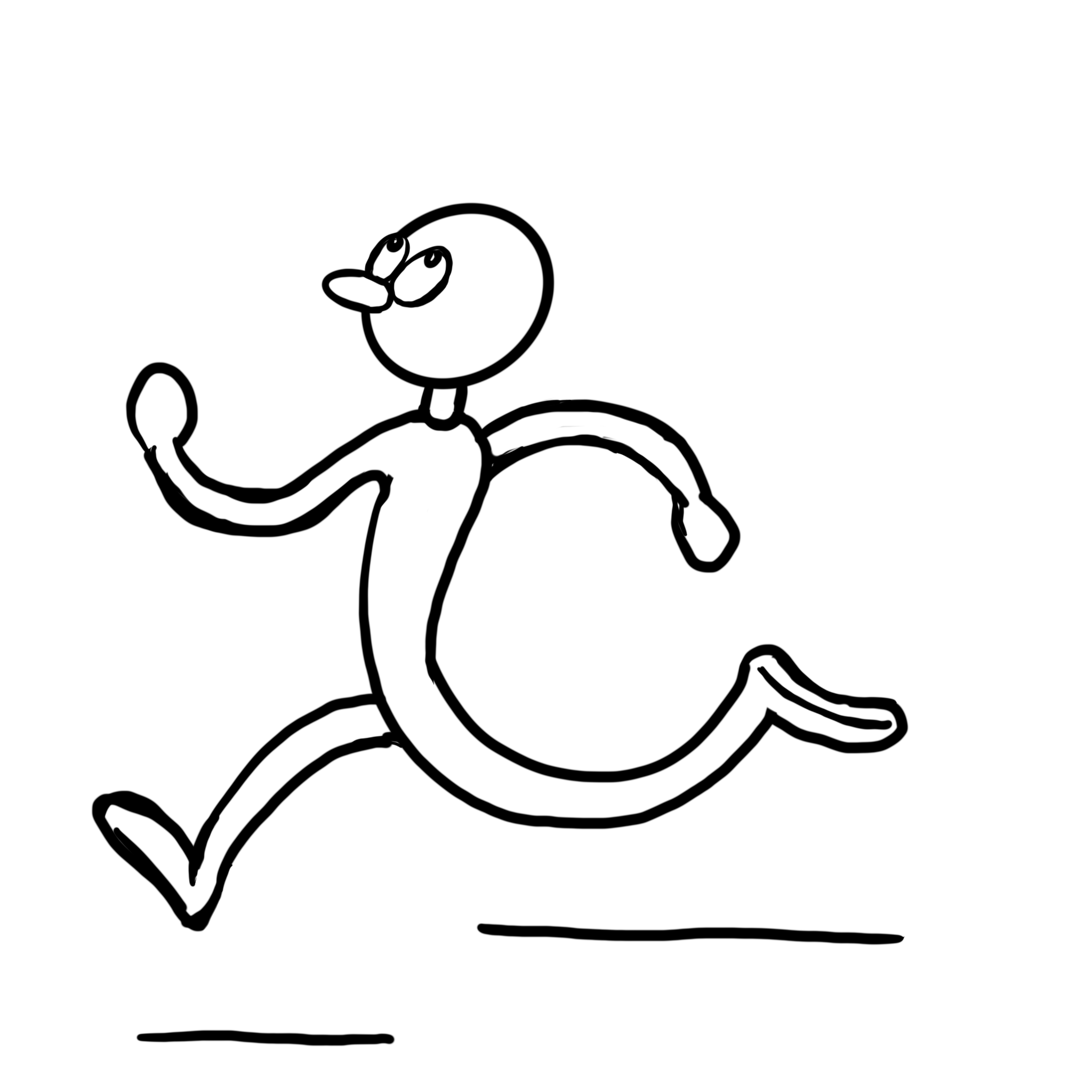}
\includegraphics[width=0.32\linewidth]{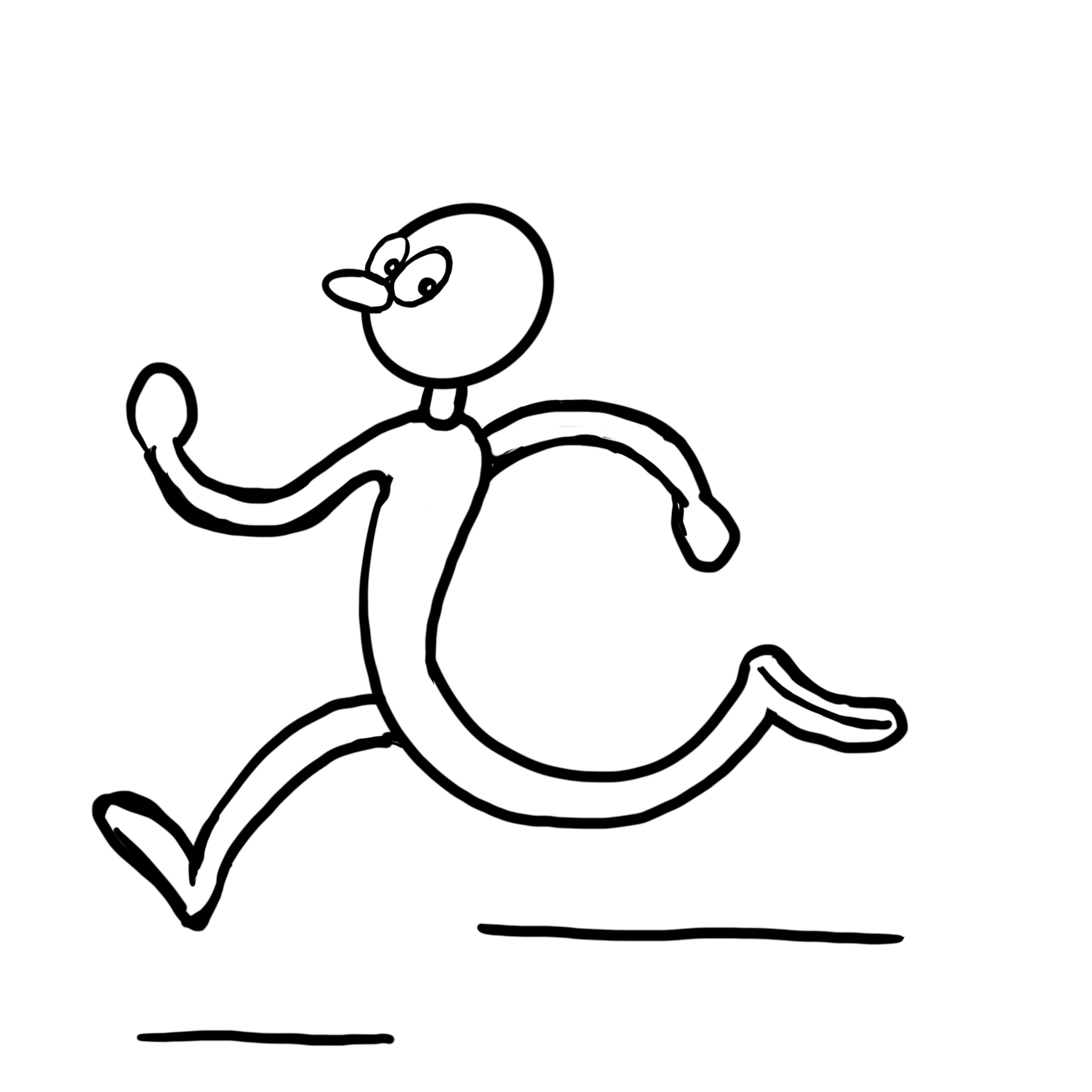}
\caption{\footnotesize Facing straight front. }
\label{1}
 \end{subfigure}
 \hfill
\begin{subfigure}[t]{0.99\linewidth}
\includegraphics[width=0.32\linewidth]{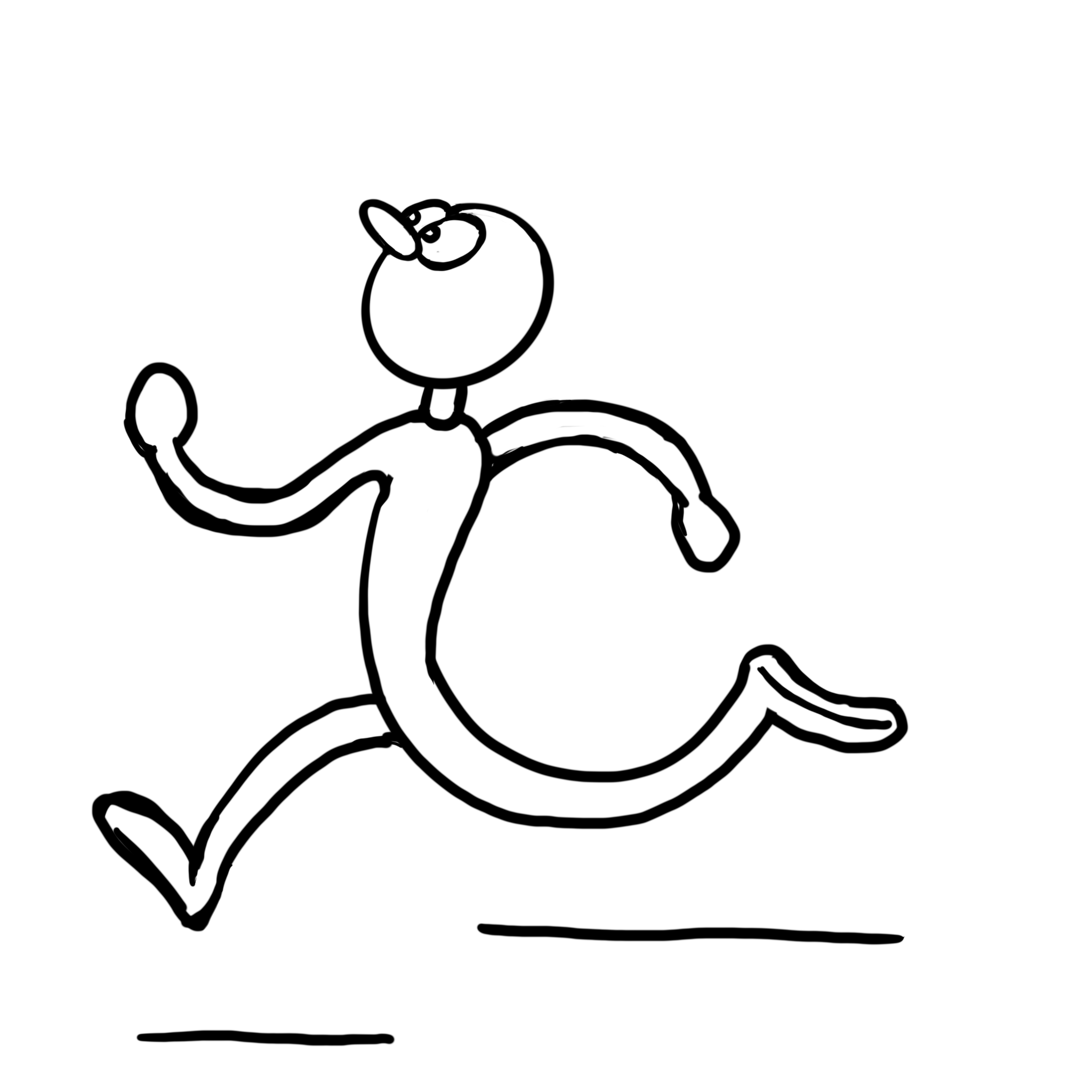}
\includegraphics[width=0.32\linewidth]{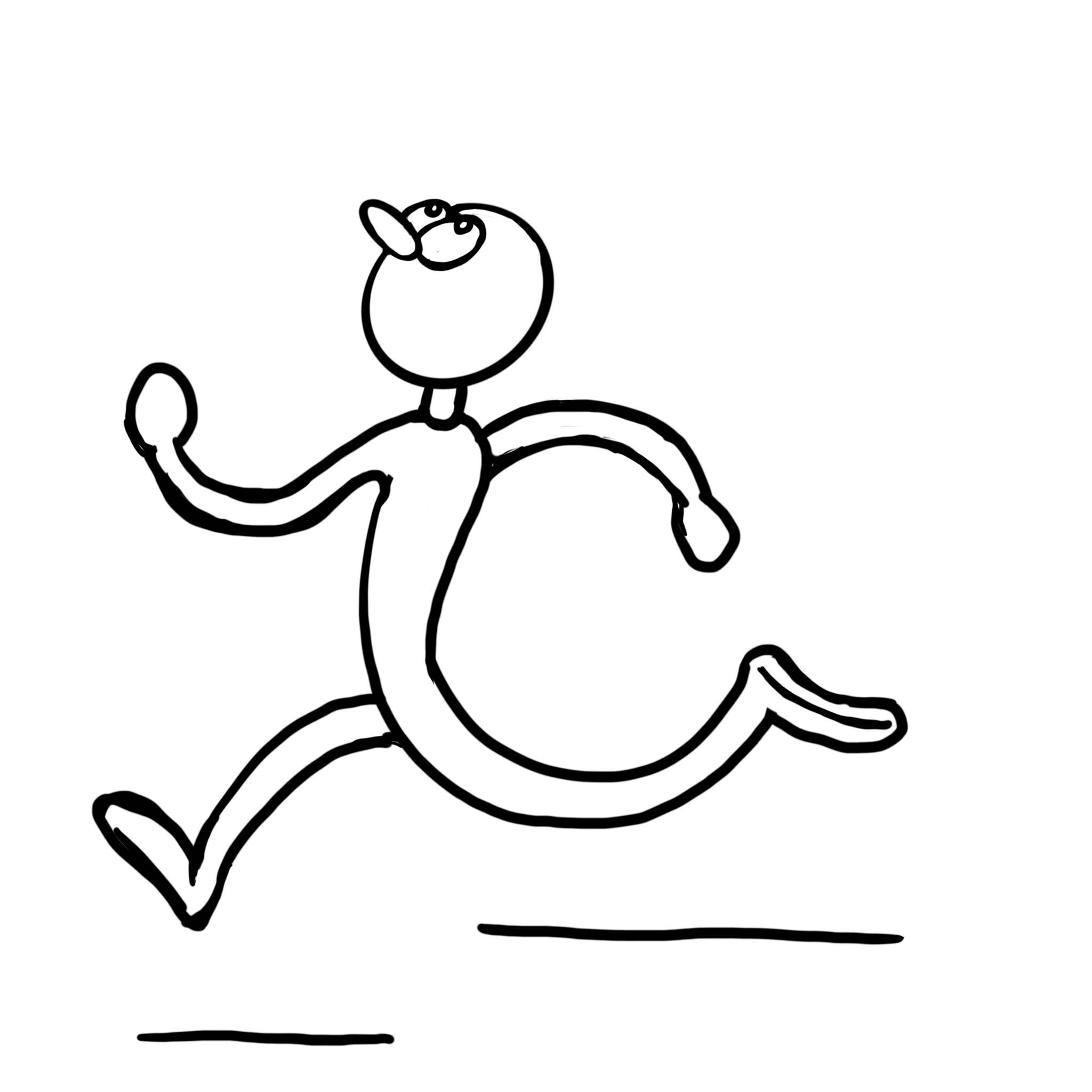}
\includegraphics[width=0.32\linewidth]{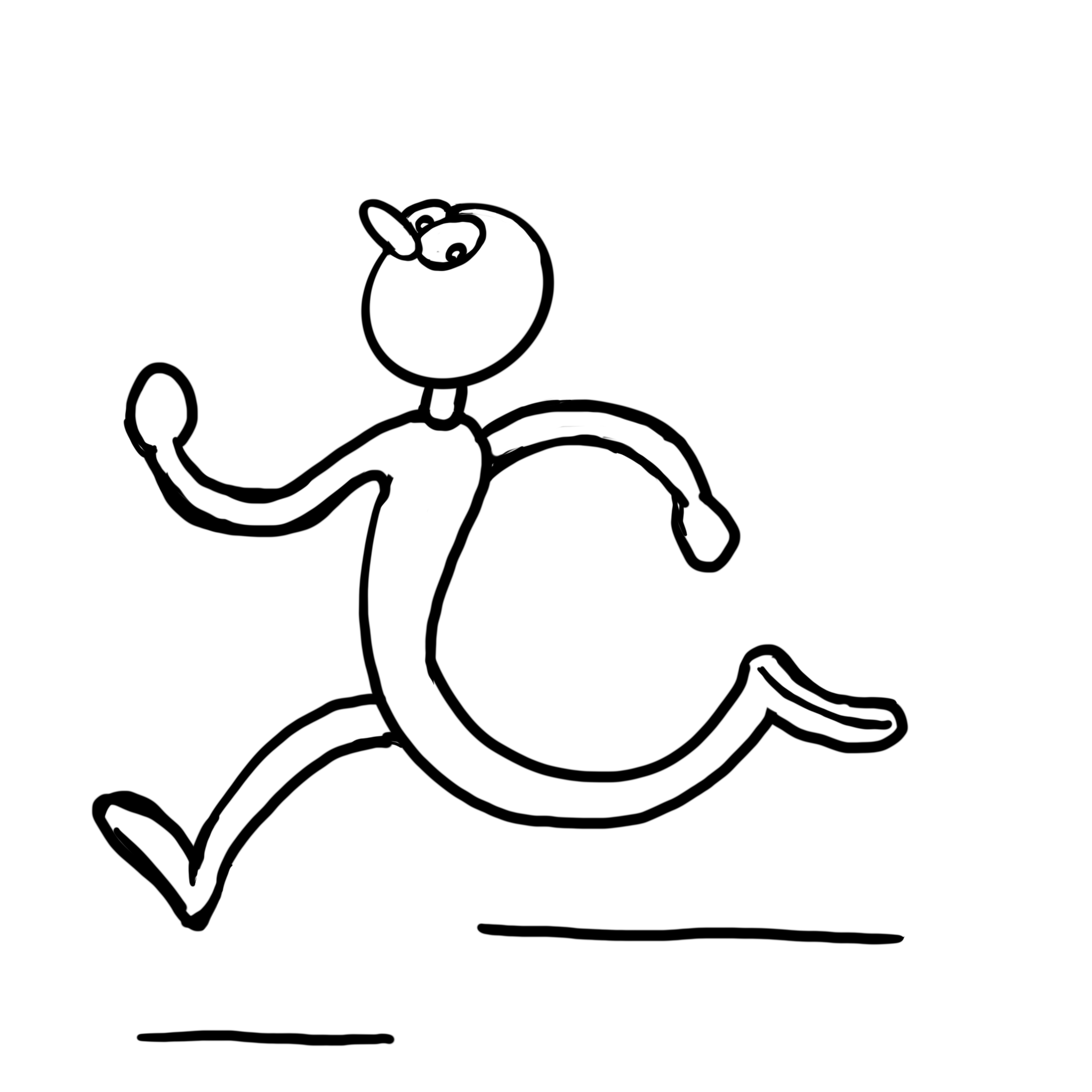}
\caption{\footnotesize Facing up and front. }
\label{fig_4/2}
 \end{subfigure}
 \hfill
\caption{The illustration set with eye-gaze direction variable.}
\label{images/4a}
\end{figure*}

\begin{figure*}[htb] 
\centering
\begin{subfigure}[t]{0.99\linewidth}
\includegraphics[width=0.32\linewidth]{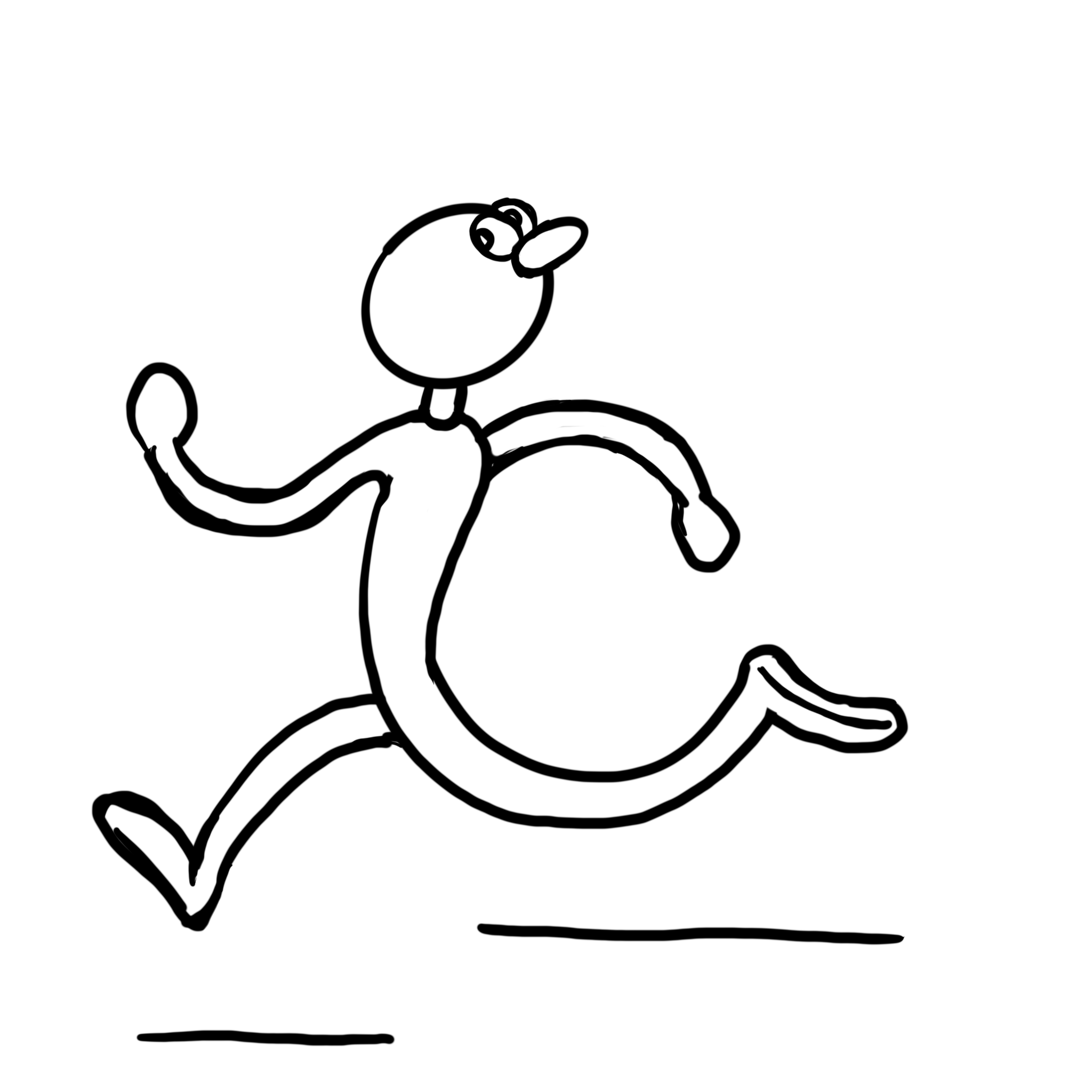}
\includegraphics[width=0.32\linewidth]{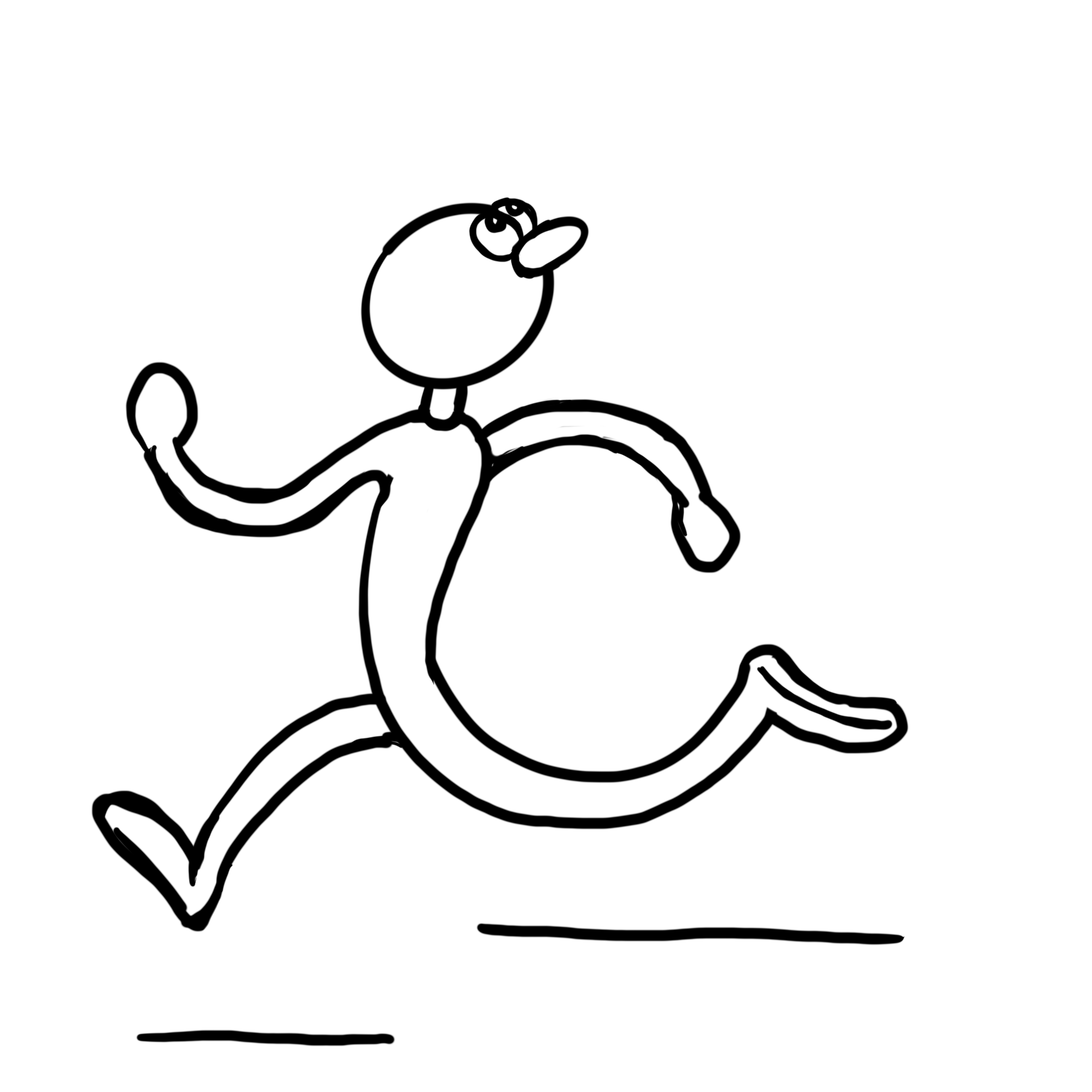}
\includegraphics[width=0.32\linewidth]{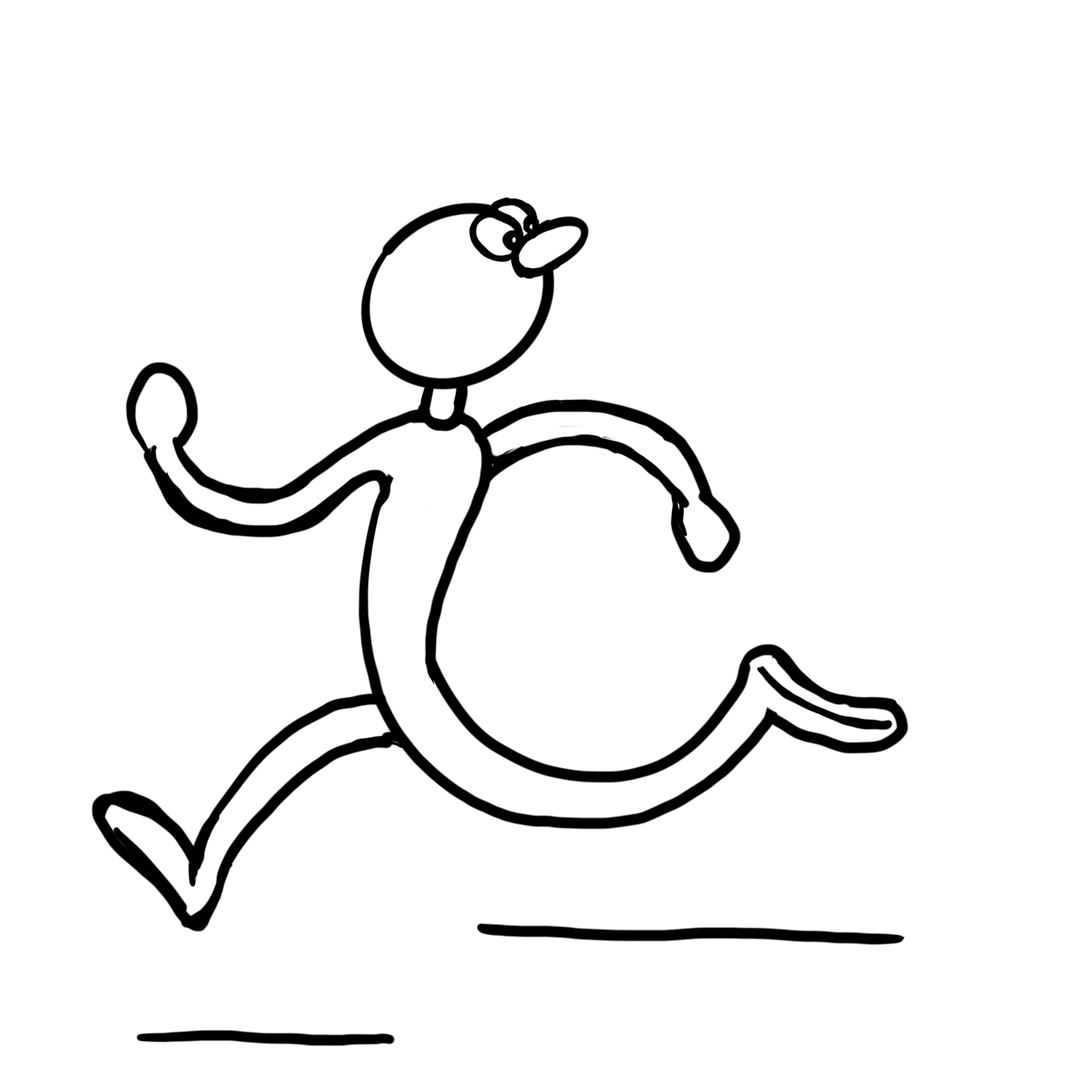}
\caption{\footnotesize Facing down and back. }
\label{fig_4/3}
 \end{subfigure}
 \hfill
\begin{subfigure}[t]{0.99\linewidth}
\includegraphics[width=0.32\linewidth]{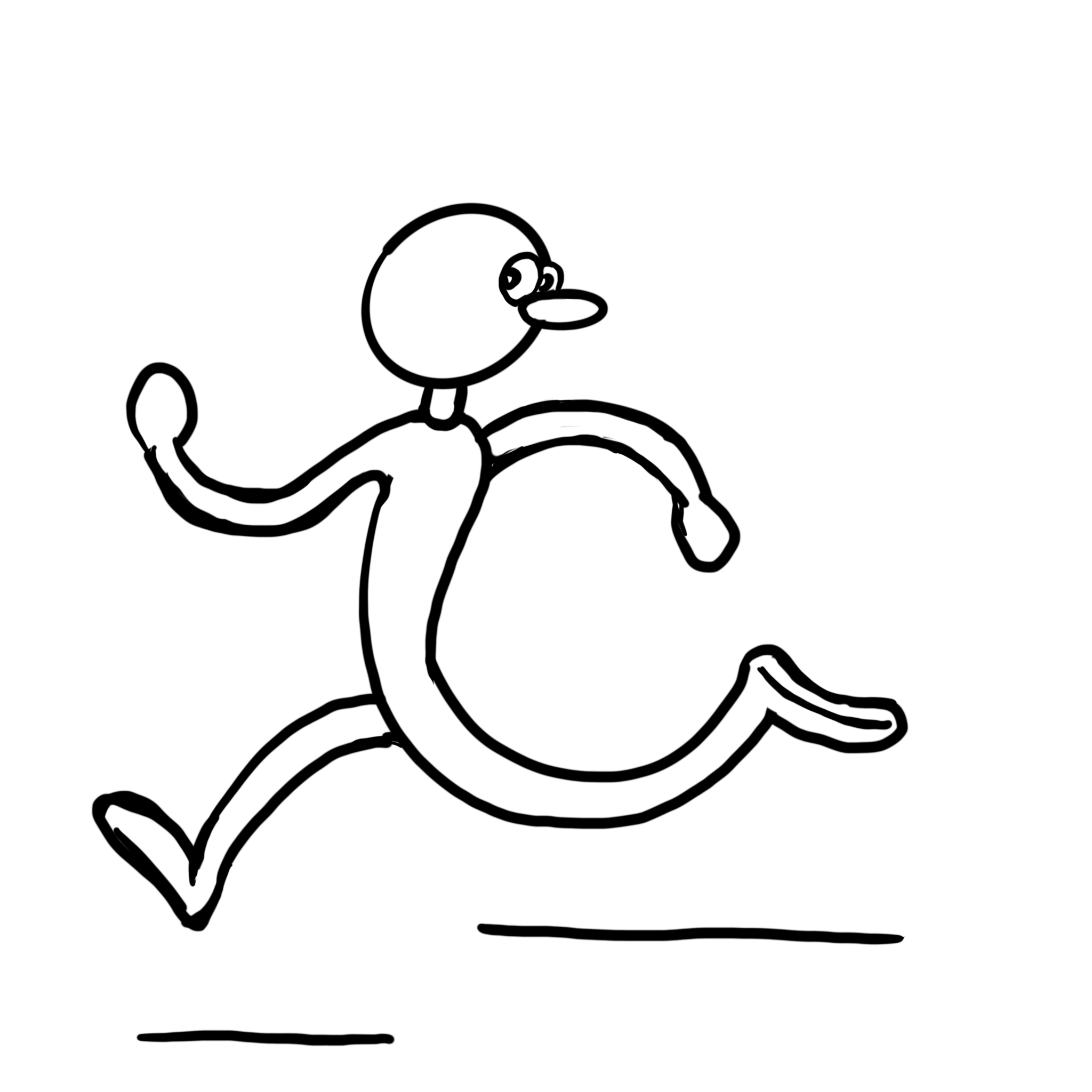}
\includegraphics[width=0.32\linewidth]{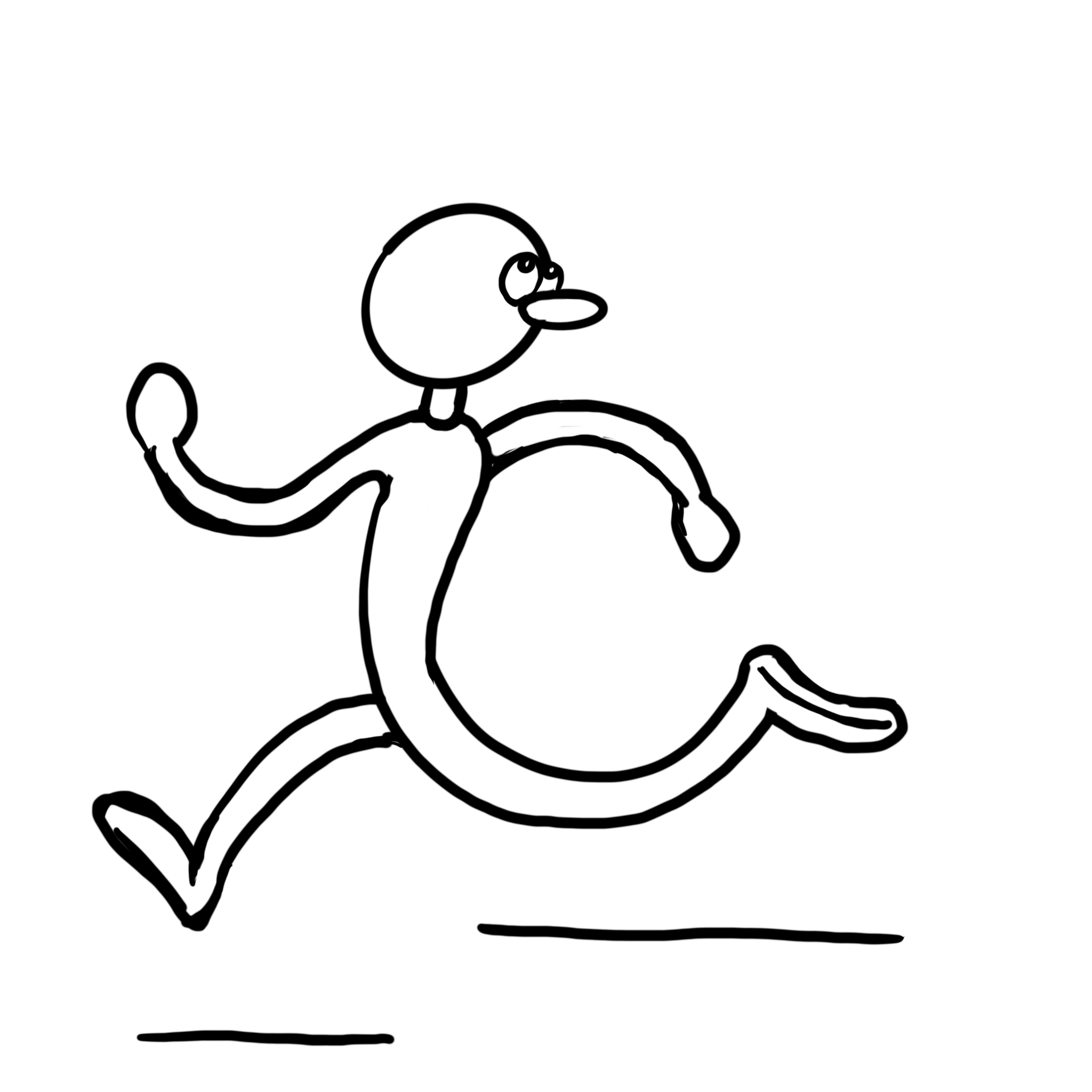}
\includegraphics[width=0.32\linewidth]{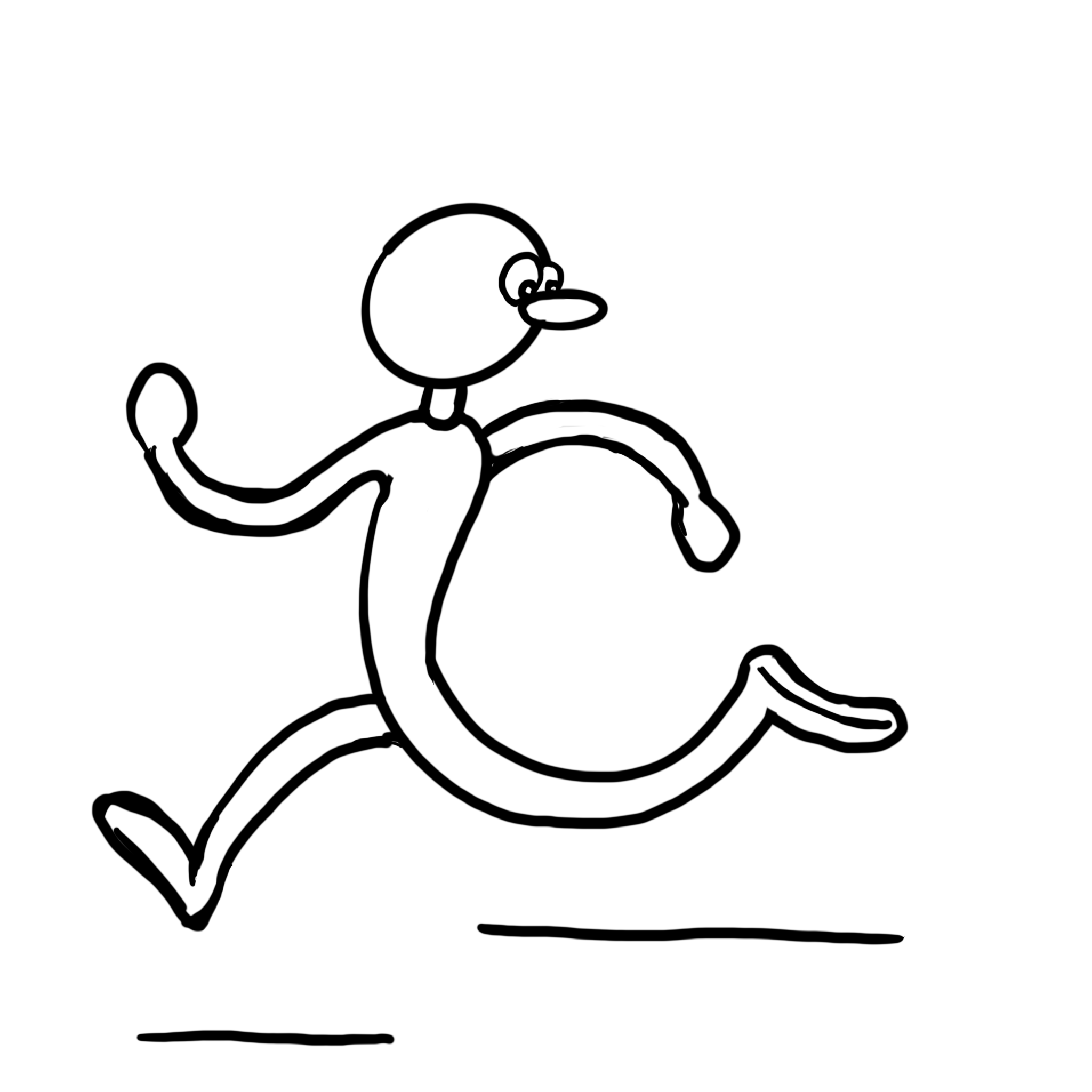}
\caption{\footnotesize Facing straight back. }
\label{fig_4/4}
 \end{subfigure}
 \hfill
\begin{subfigure}[t]{0.99\linewidth}
\includegraphics[width=0.32\linewidth]{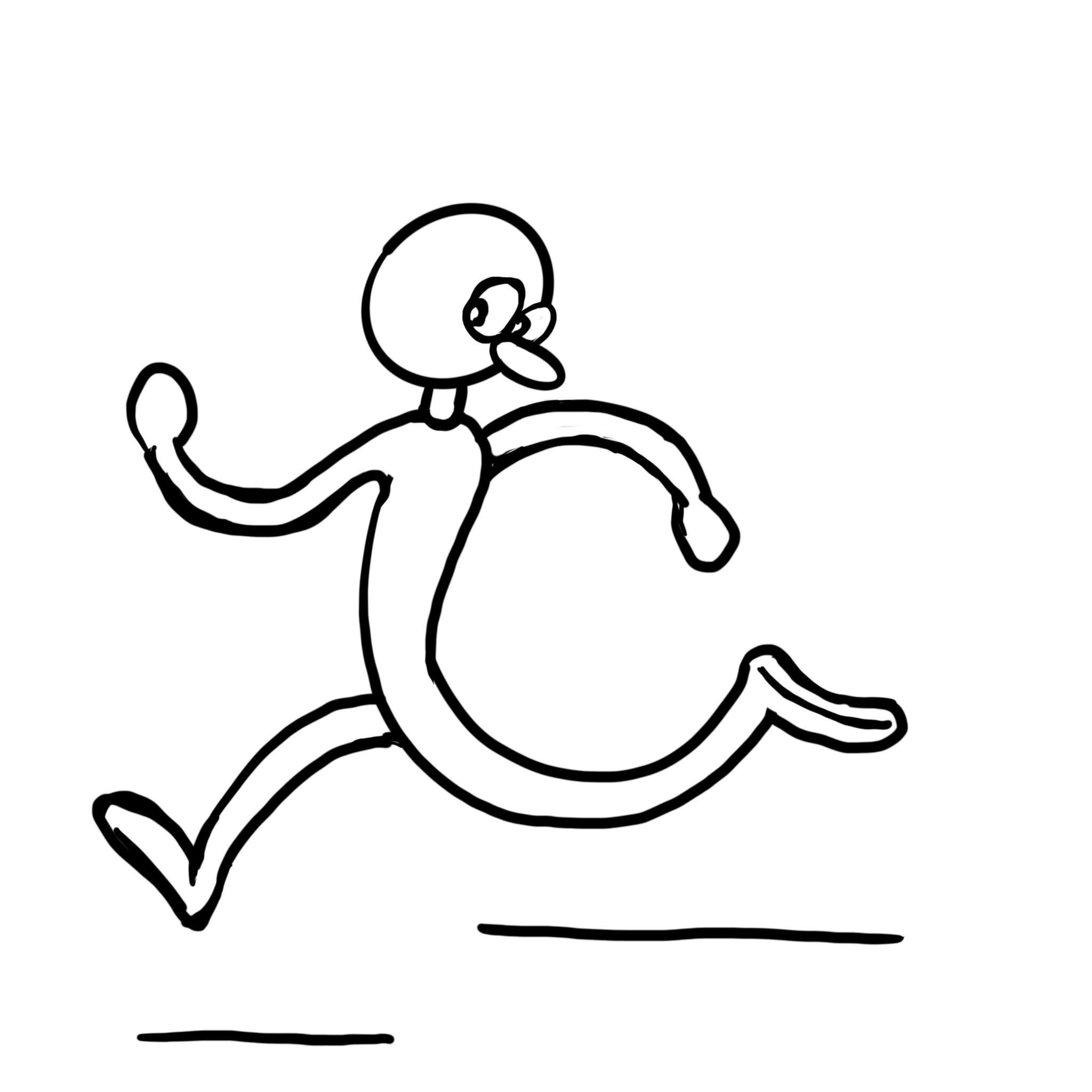}
\includegraphics[width=0.32\linewidth]{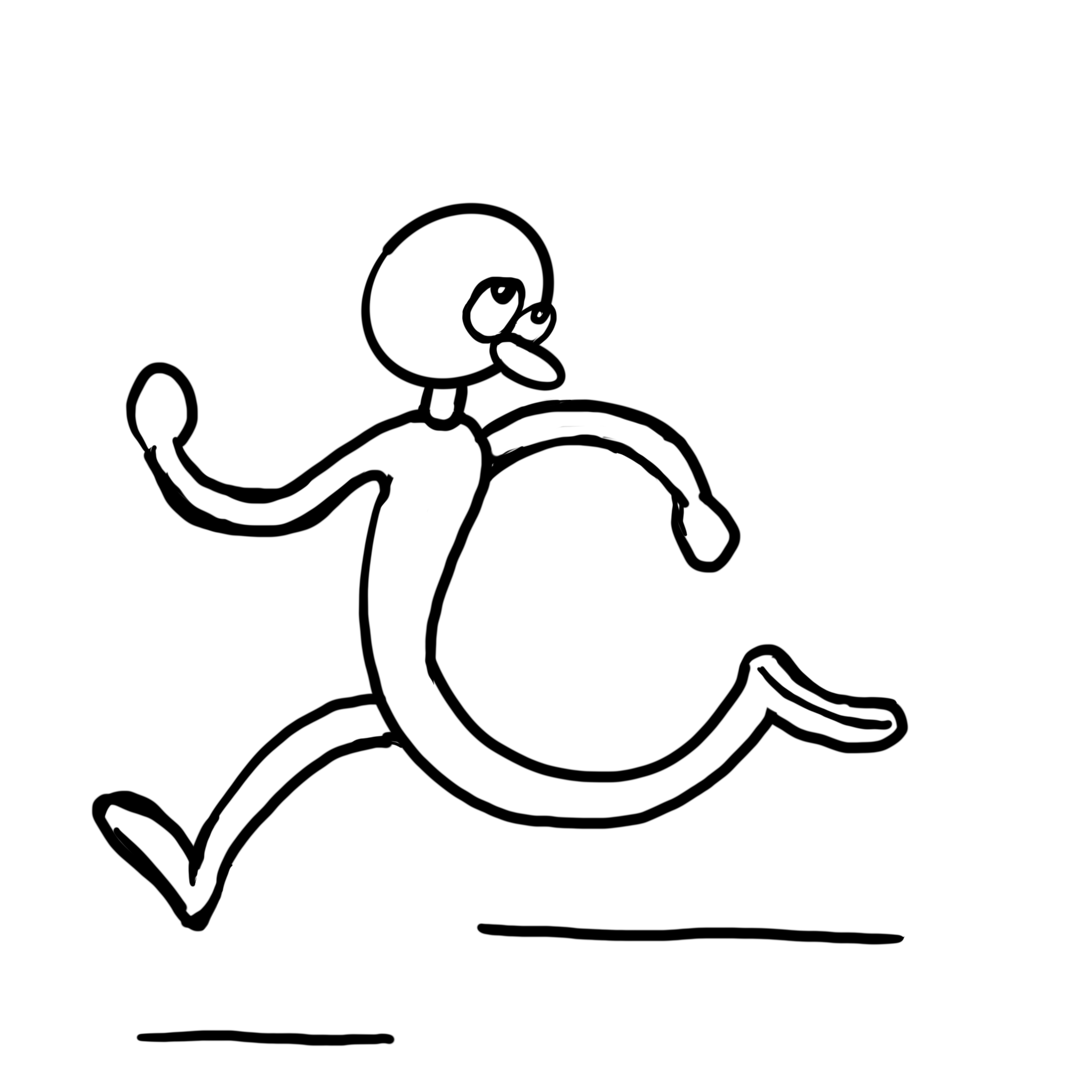}
\includegraphics[width=0.32\linewidth]{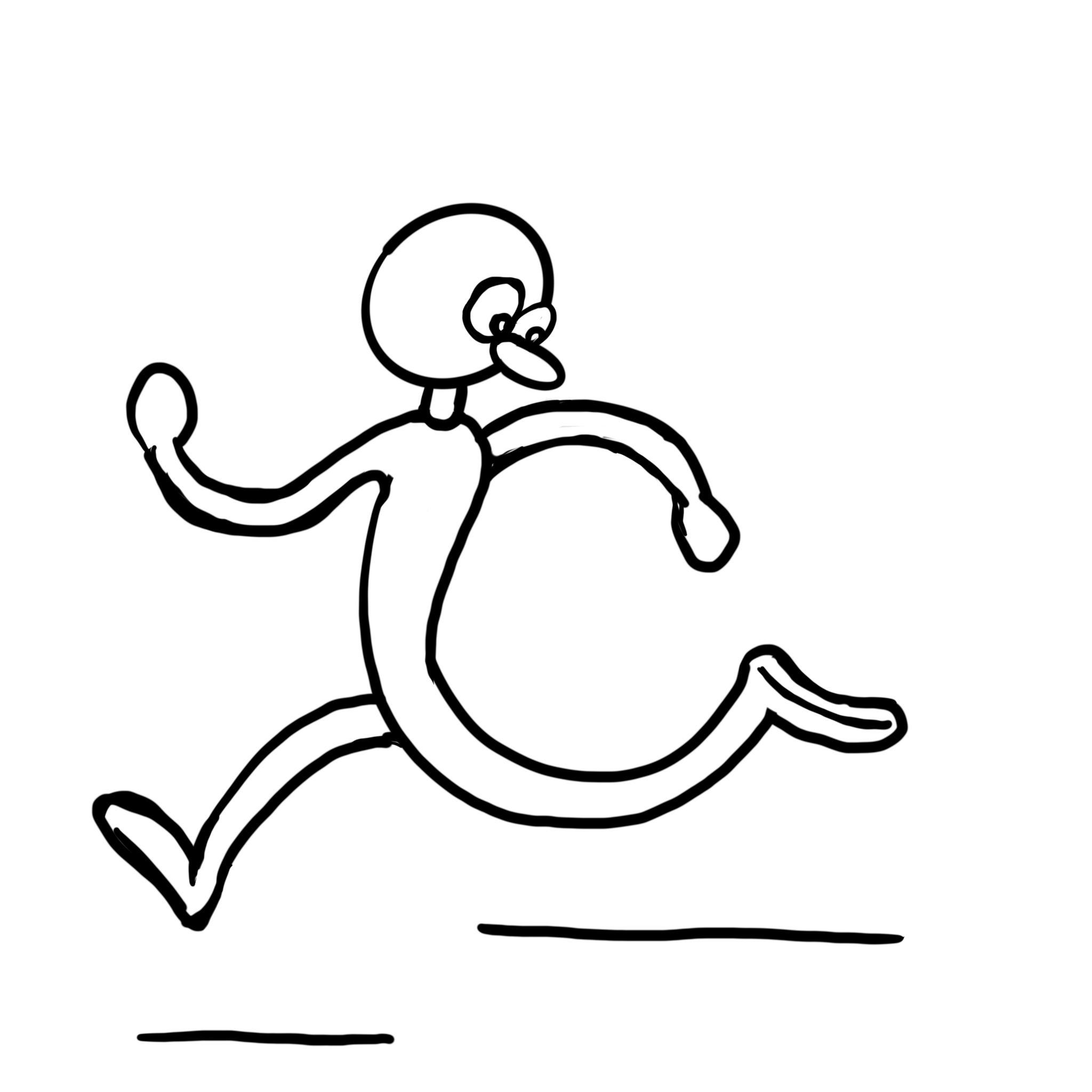}
\caption{\footnotesize Facing up and back. }
\label{fig_4/5}
 \end{subfigure}
 \hfill
\caption{The illustration set with eye-gaze direction variable.}
\label{images/4b}
\end{figure*}

The complete set of data we used for our study is shown in Figures~\ref{images/4a} and ~\ref{images/4b}. We focus on investigating the effect of the gaze direction of a running man on the effective context perceived by the viewer. We have shown these images to independent volunteers and asked them to describe the event that took place in the scene and the emotion associated with the figure. From the results, we have observed that the change in the direction of the nose and eyes creates different contexts and emotions even though the rest of the body is the same.
We have also considered an additional illustration set, adding gaze direction as a second variable to see if that affects the context as well; however, the limitations of our user study did not allow us to investigate that variable (see Figures~\ref{images/4a} and ~\ref{images/4b}).

\section{Related Work}

One of the main influences in the creation of different expressions in illustrations comes from Ollie Johnston and Frank Thomas,  two of Walt Disney's leading teams of animators known as the Nine Old Men \cite{johnston1981illusion}. They also presented the animation principles developed by Disney Studios to the general public.  The animation principles were later adapted to computer graphics applications by another legendary animator John Lasseter \cite{lasseter1998principles}. Another influential animator in understanding and creating subtle expressions by sharing his vast practical knowledge is Preston Blair \cite{blair1995cartoon}. Cartoonists such as Scott McCloud \cite{mccloud1993understanding,mccloud2006making} and Will Eisner \cite{eisner2008comics, eisner2008graphic} also contributed greatly to the understanding of this problem. We can also list Hakan Celik and Ergun Akleman who showed the importance of some subtle expressive cues \cite{celik2011on,akleman2020}. 

Despite relatively few publications on the art side, there is a large body of literature on the intersection of human-computer interaction and psychology to analyze and understand affective states \cite{picard2000affective, picard2003affective, tao2005affective}. Harmon et al. classify affective states in three main dimensions: valence, arousal, and motivational intensity \cite{harmon2011attitudes,harmon2013does,harmon2013approach,gable2013does}. We observe that subtle cues can also be important in analyzing these dimensions effectively.

\section{Methodology}

The key part of our methodology is to create a simple set of sketch-based illustrations that can be easily shown to participants in the user study. We have three hypotheses that we want to investigate with the user study:
\begin{enumerate}
\item Looking behind while running will invoke a context happening out of the ordinary compared to looking ahead while running illustrations in Question-1.
\item Looking behind while running will trigger more negative emotion descriptions compared to looking ahead while running illustrations in Question 2.
\item The intensity of the emotion will increase when the illustrated man is looking up or down, compared to the man looking straight in Question 3.
\end{enumerate}

For the user study, we collected data using Qualtrics in Turkish. The volunteer group consists of 54 random individuals from different age groups, occupations, and gender preferences. The answers to these questions are taken from the online platform without direct contact with the participants. No personal data is collected during this process.

It might be worth noting that they all share the same cultural heritage as the experiment was conducted only by Turkish citizens. Illustrations are presented in a random order to counter-balance the order effect. For every illustration, we asked participants to answer three open-ended questions: 
\begin{enumerate}
\item What do you think that is going on in this illustration? 
\item How does the illustrated man feel at the moment?
\item Can you rate the intensity of this emotion from 1 to 10? (1 means "almost none" and 10 means "overwhelmingly").
\end{enumerate}

We refrain from offering choices to the participants to avoid inducing bias. We considered the position "facing front" as our baseline and compared our results from different positions with respect to that position. If a user identified an emotion but did not include the intensity, we did not include that entry in our result analysis. Users who did not complete all the questions but answered all the questions related to a single illustration were included in the results of that illustration. 

\section{Results}

To analyze our results, we have used one of the methods of qualitative analysis called Grounded Theory. We counted the number of participants who have used "running" and "fleeing" while describing the illustrated figure. A chi-square analysis of X2(2) = 13,64, p<0.01 showed significant results for using the word 'running' to describe a man facing the front compared to describing a man facing the back (see Table 1). However, the results were not significant for the "fleeing" word count (see Table 2). 

To describe the man facing down and front, 21 descriptions made out of 54 in total gave details about the reasons for looking to that specific position. Similarly, 19 descriptions made for the man facing up and front explained the reasons for that direction. On the other hand, only five descriptions out of 54 descriptions made for the facing-front illustration tried to explain the reasons for looking that way. 

The intensity comparison showed that the back conditions produce significantly higher intensity for emotions of fear, concern, curiosity, and excitation compared to the front. Furthermore, facing up yielded high intensity scores for emotions such as happiness, curiosity, and excitation, while facing down yielded high scores for fear, concern, ambition, and anxiety. Lastly, the difference in the intensity of emotions between facing up and facing ahead is relatively high; people tend to rate emotions more intensely when the character is facing up. 

 \begin{figure}
    \centering
    \begin{subfigure}[t]{1\linewidth}
\includegraphics[width=0.99\linewidth]{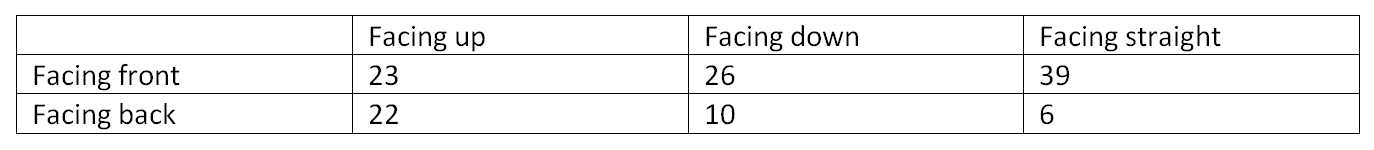}
\caption{This table shows the number of participants who used the word "running" to describe the illustrated man.}
\label{Table/1}
    \end{subfigure}
    \begin{subfigure}[t]{1\linewidth}
\includegraphics[width=0.99\linewidth]{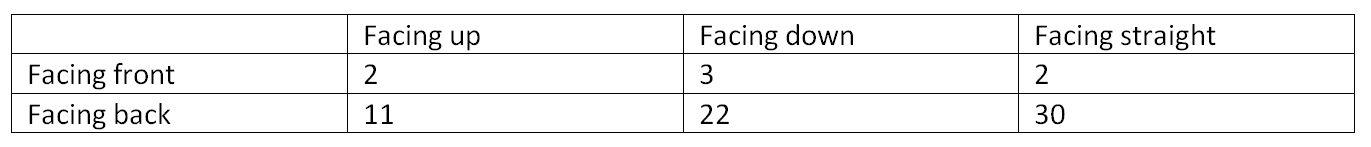}
\caption{This table shows the number of participants who used the word "to flee" to describe the illustrated man.}
\label{Table/2}
    \end{subfigure}
\end{figure}

\section{Discussion}

The immediate result to report is the change of words while describing the action when the eye and nose orientations change. Tables 1 and 2 show the usage of the words "running" and "fleeing", respectively. We see that the word running is used more often when the figure is facing the front, whereas the word "fleeing" is used when the figure is facing backwards. 

Some participants reported that they have focused on the absence of some features in addition to the features existing in the illustrations. For example, one participant reported: "I had thought the guy was running a marathon, but then I realized that there was no sweat drawn. Thus, I have changed my answer." Some participants perceived the illustrations as scenes following each other and responded accordingly. For example, one participant responded "Somebody threw a rock to the guy" while describing the man facing up and back, then followed with the description "The rock did not hit the guy, fell next to him." illustration facing front and down. 

\begin{figure}
\centering
    \begin{subfigure}[t]{0.45\linewidth}
\includegraphics[width=0.99\linewidth]{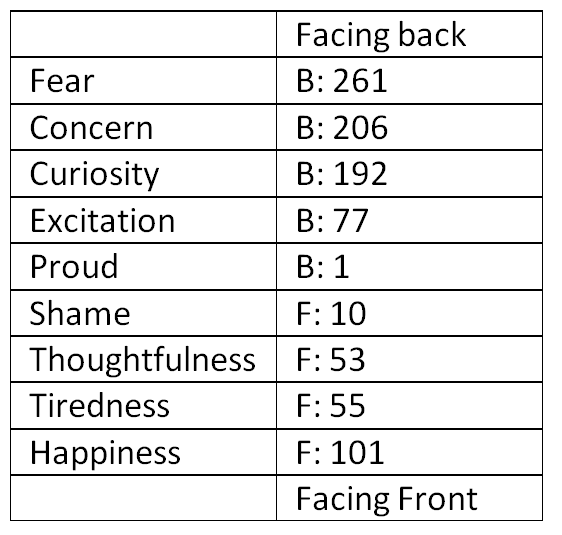}
\caption{Facing Back and Front.}
\label{Table/3}
    \end{subfigure}
    \hfill
    \begin{subfigure}[t]{0.45\linewidth}
\includegraphics[width=0.99\linewidth]{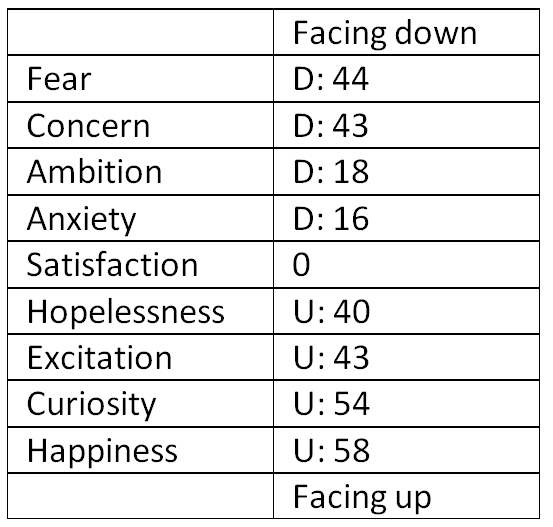}
\caption{Facing Up and Down.}
\label{Table/4}
    \end{subfigure}
\captionsetup{justification=centering}
\caption{These two tables show the intensity rating comparison for different emotions. Table 3 compares the highest emotion ratings for facing back and front conditions. B: facing back condition intensity rating. F: Facing front condition intensity rating. Table 4 compares the highest emotion ratings for facing up and down conditions. U: facing up condition intensity rating. D: Facing down condition intensity rating.  .}
\label{Table/3-4}
\end{figure}

We have seen that when the figure is facing down, both the descriptions and the emotions suggested for those scenes had negative valence. In contrast, positive valence emotions were attributed to the figure facing up with one exception of hopelessness. To measure the significance of the associated emotions, we use an intensity value. We acquire the intensity for each emotion for that illustration by summing the intensity values assigned by the participants. It should be noted that a single participant can assign a maximum of three emotions to a given illustration. We then look at the difference between the intensity of the same emotion in different poses for comparison. If an emotion is named in one pose and not in the other, its intensity value is considered 0 for the figure that lacks the emotion for this analysis. Table 3 shows the results by comparing the figure facing forward and facing backward. As we can see from the table, the words "fear, concern, curiosity, excitation" are more associated with facing backward than forward, while "thoughtful, tired, happiness" is more associated with facing forward than backward. Although no conventional methods to express emotions (such as eyebrows and mouth shape) were used in the illustration, the positioning of the figure itself was enough to attribute these emotions to the figures. This means that our internal states that are shaped by cultural and personal preoccupations are affected by small details in shaping a context. Moreover, this study shows that creating and describing a context is more complex than it seems, since it is a compilation of many fine details that may be obscure initially. 

We believe that future studies will benefit from a more detailed analysis of the context and the elements that create the context. A proper understanding of the context and its implementation will greatly improve the performance and user experience in intelligent user interfaces.

\section{Conclusion and Future Work}

In this work, we have developed a method to test hypotheses about the impact of subtle expressive cues on the perception of human affects. Using this method, we show that gaze direction in a running person can change perceived human affects. Although our method can be used to formally demonstrate the impact of other subtle cues, we think that a formal scientific user study can be overkill for most of these effects. Our study demonstrates that there is a need for a series of short publications from animators, illustrators, and cartoonists to explain the impact of some of these affects using their artworks as examples. 

\bibliographystyle{unsrtnat}
\bibliography{references}

\begin{figure*}[htb] 
\centering
\includegraphics[width=0.32\linewidth]{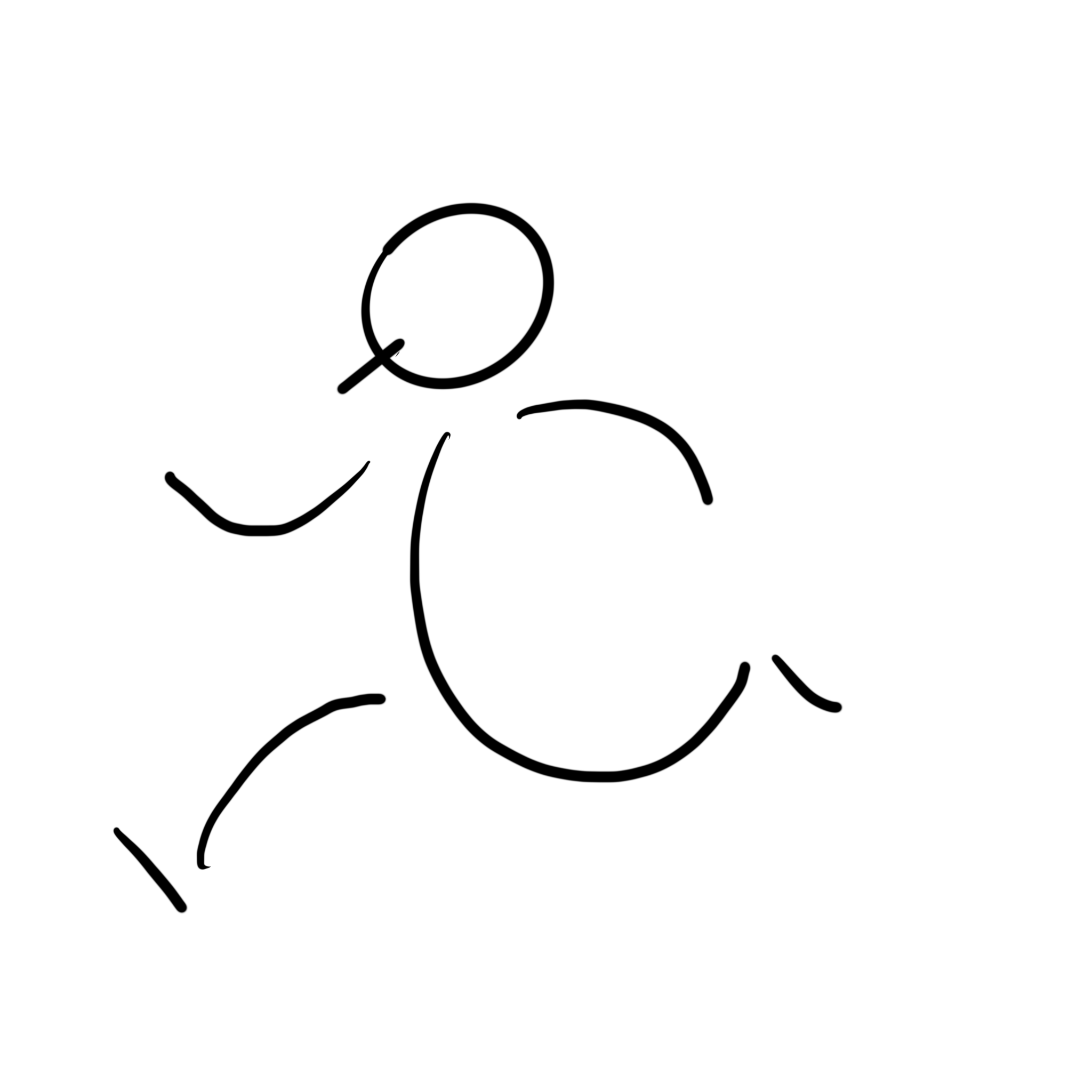}
\includegraphics[width=0.32\linewidth]{0/01}
\includegraphics[width=0.32\linewidth]{0/02}
\includegraphics[width=0.32\linewidth]{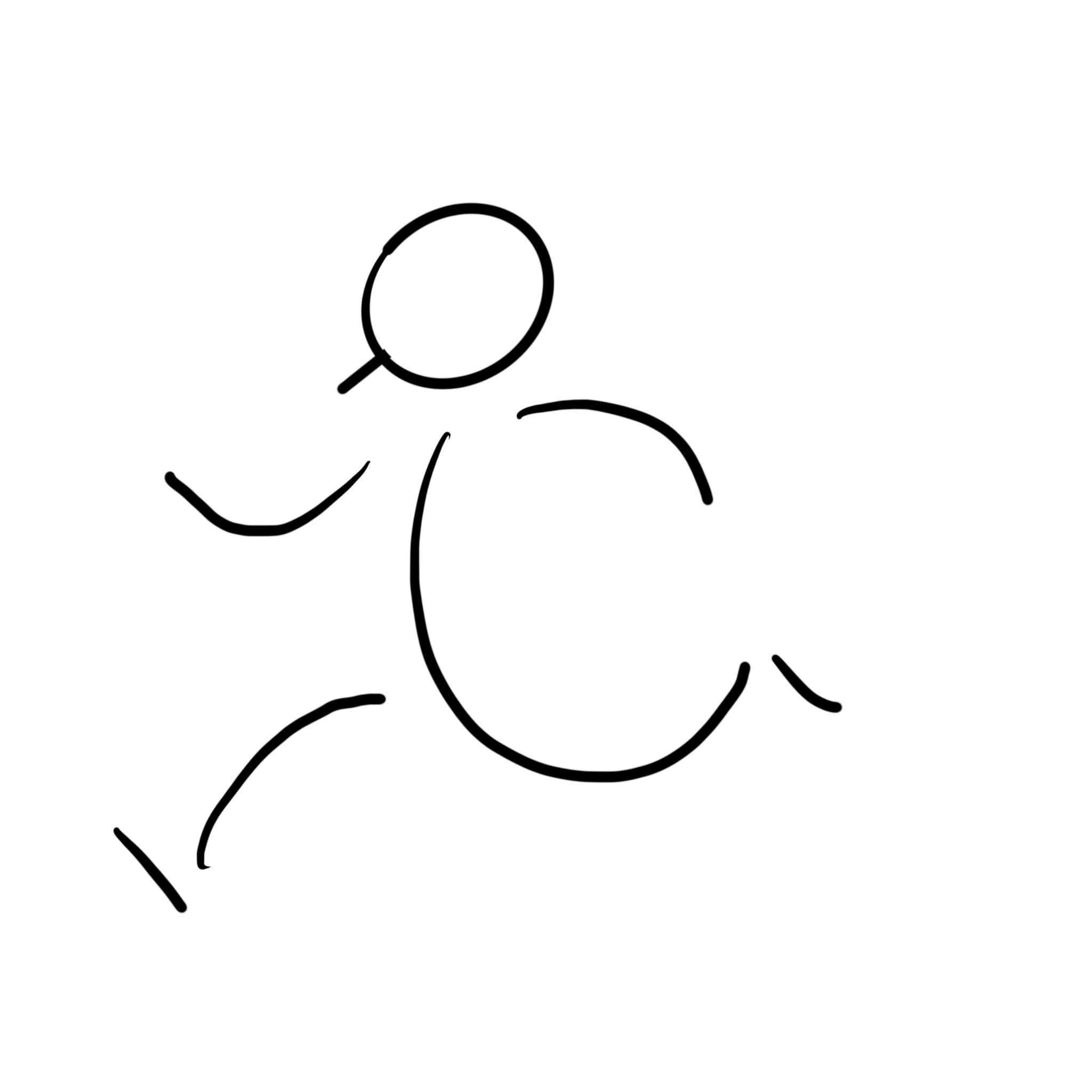}
\includegraphics[width=0.32\linewidth]{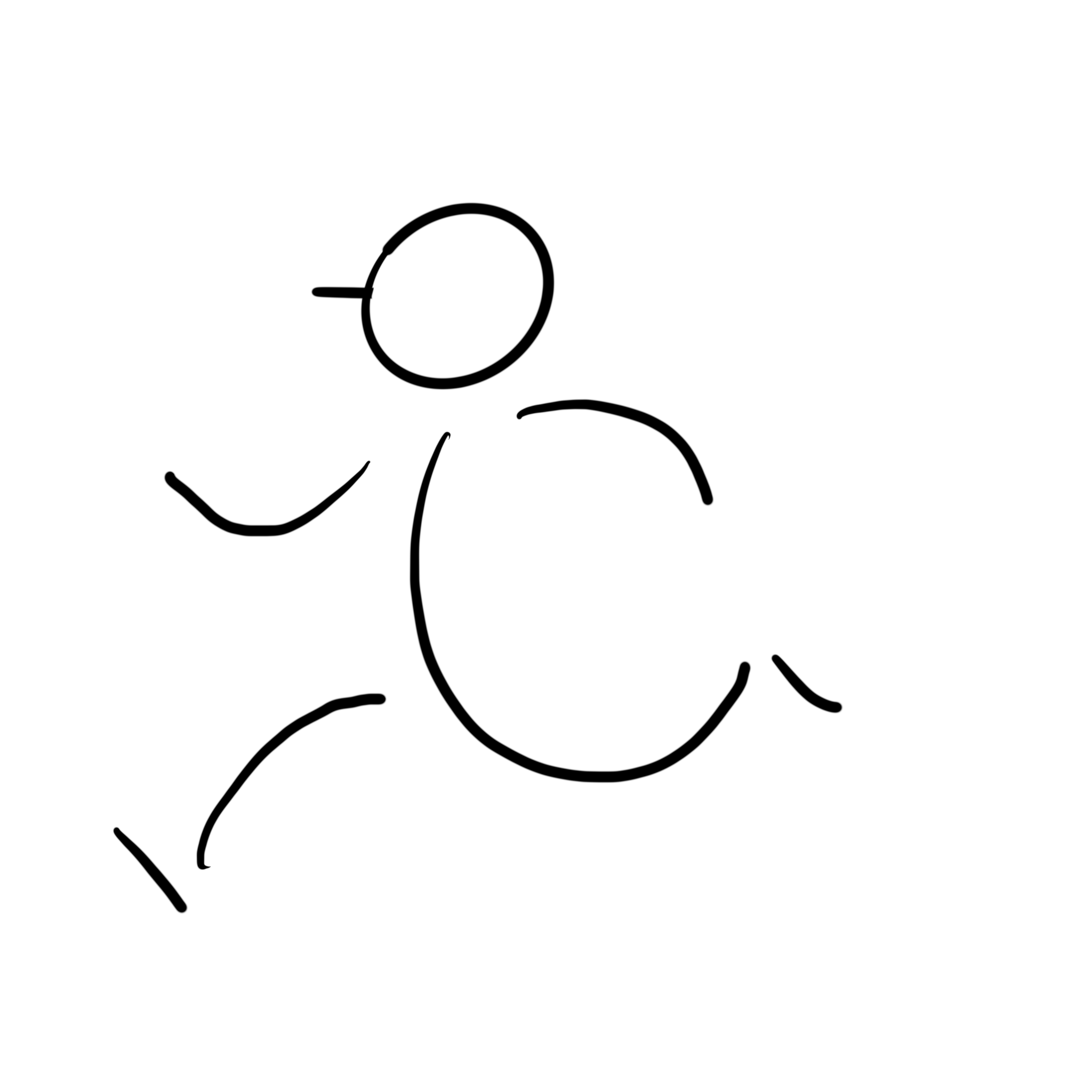}
\includegraphics[width=0.32\linewidth]{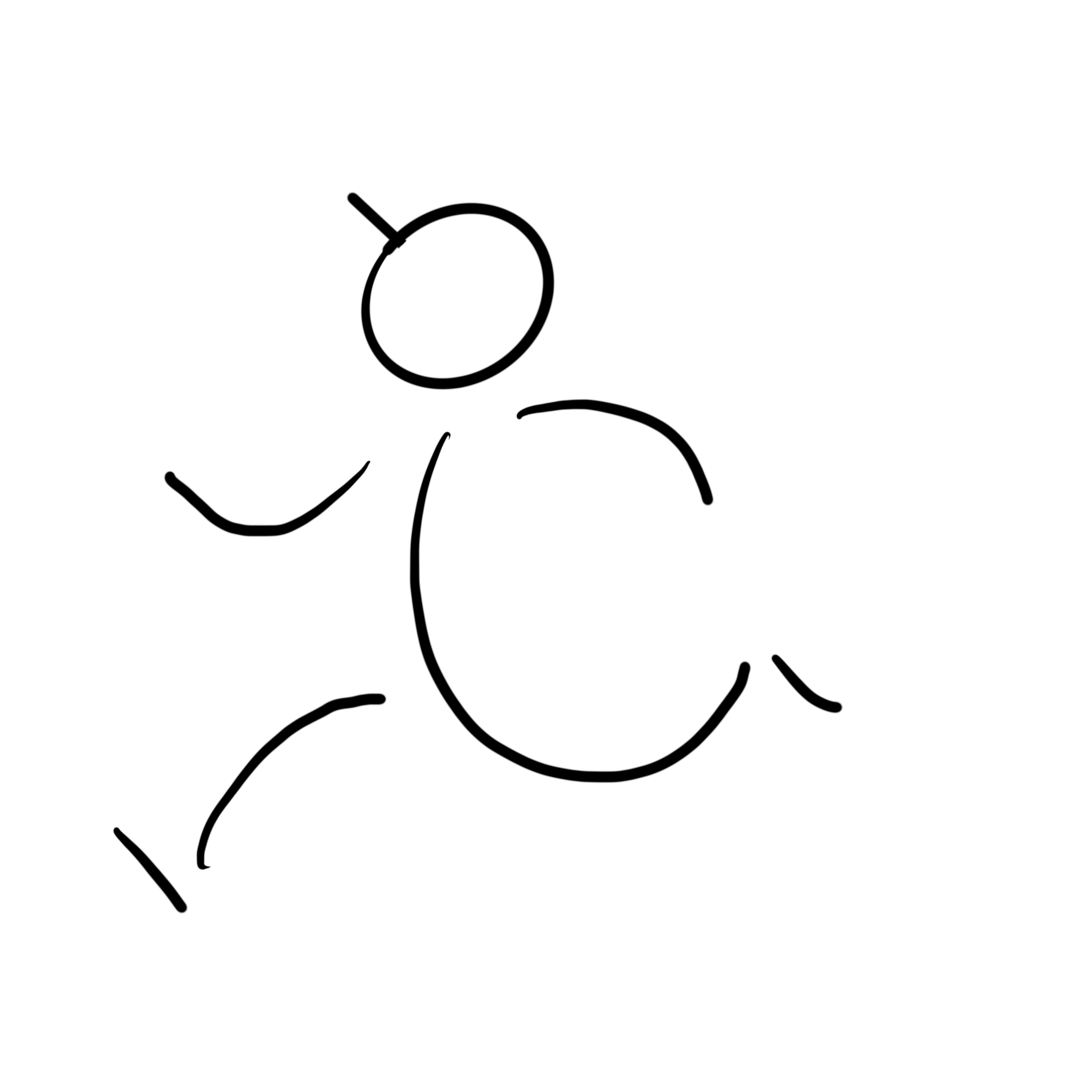}
\caption{Set-0: wire-person running.}
\label{images/0}
\end{figure*}

\begin{figure*}[htb] 
\centering
\includegraphics[width=0.32\linewidth]{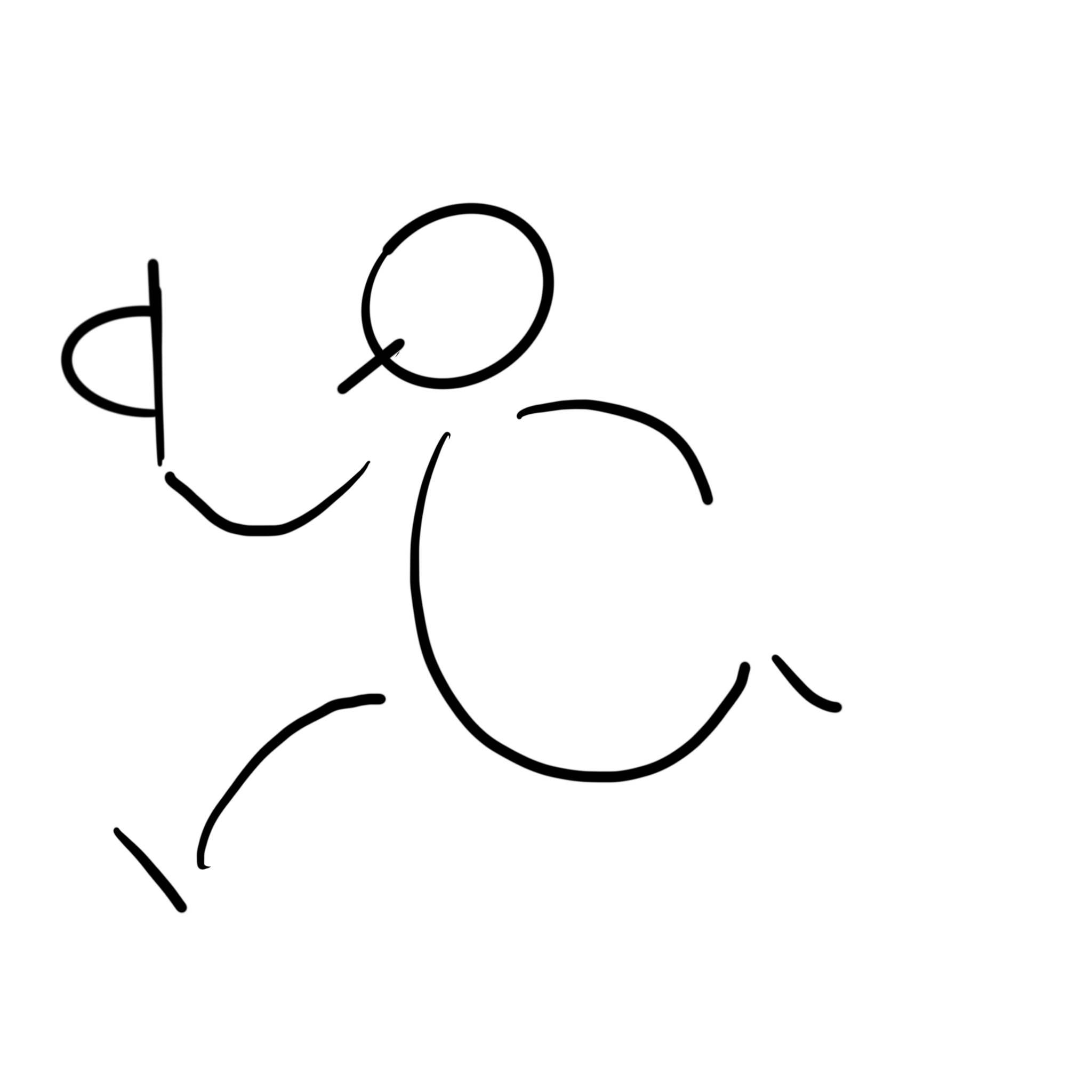}
\includegraphics[width=0.32\linewidth]{1/01}
\includegraphics[width=0.32\linewidth]{1/02}
\includegraphics[width=0.32\linewidth]{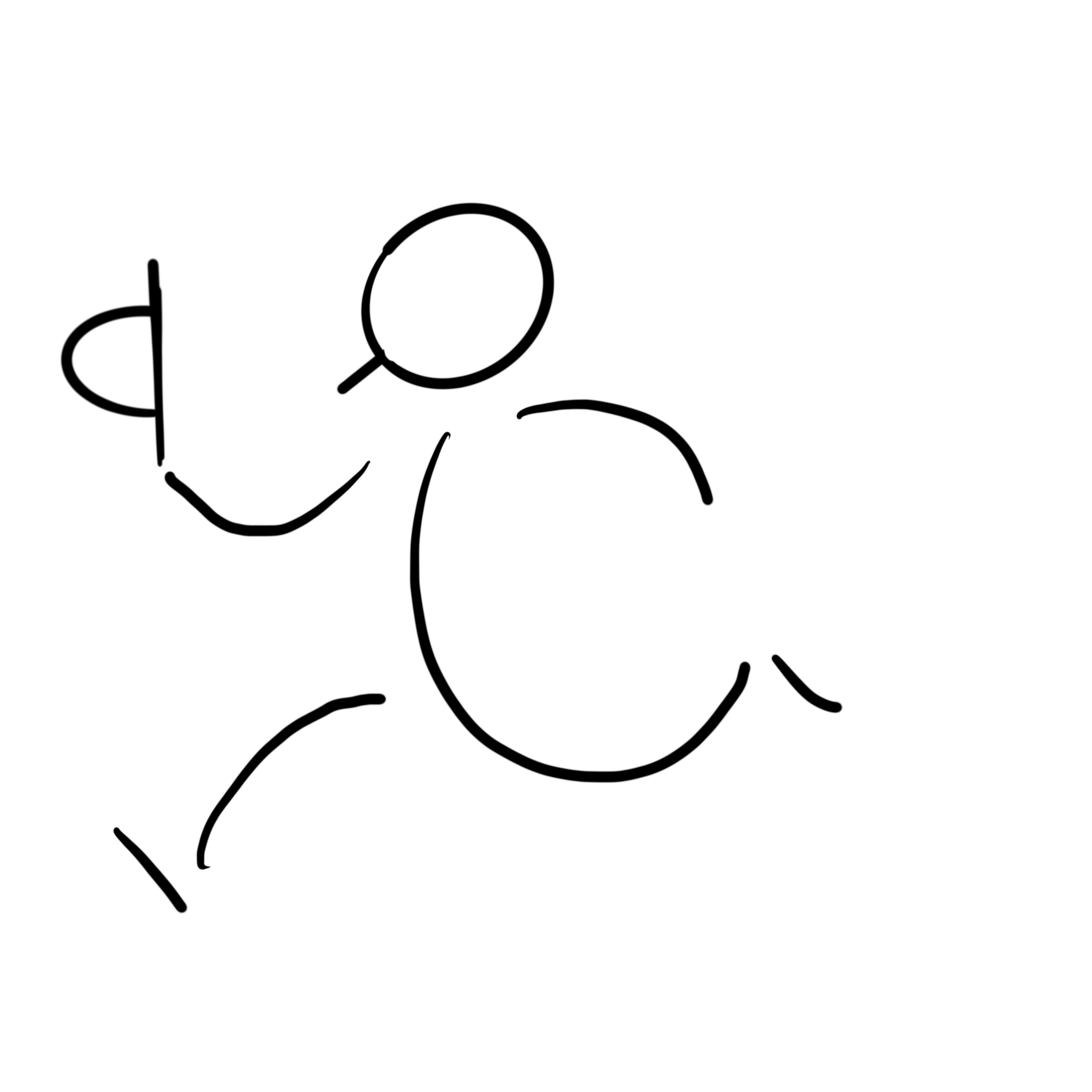}
\includegraphics[width=0.32\linewidth]{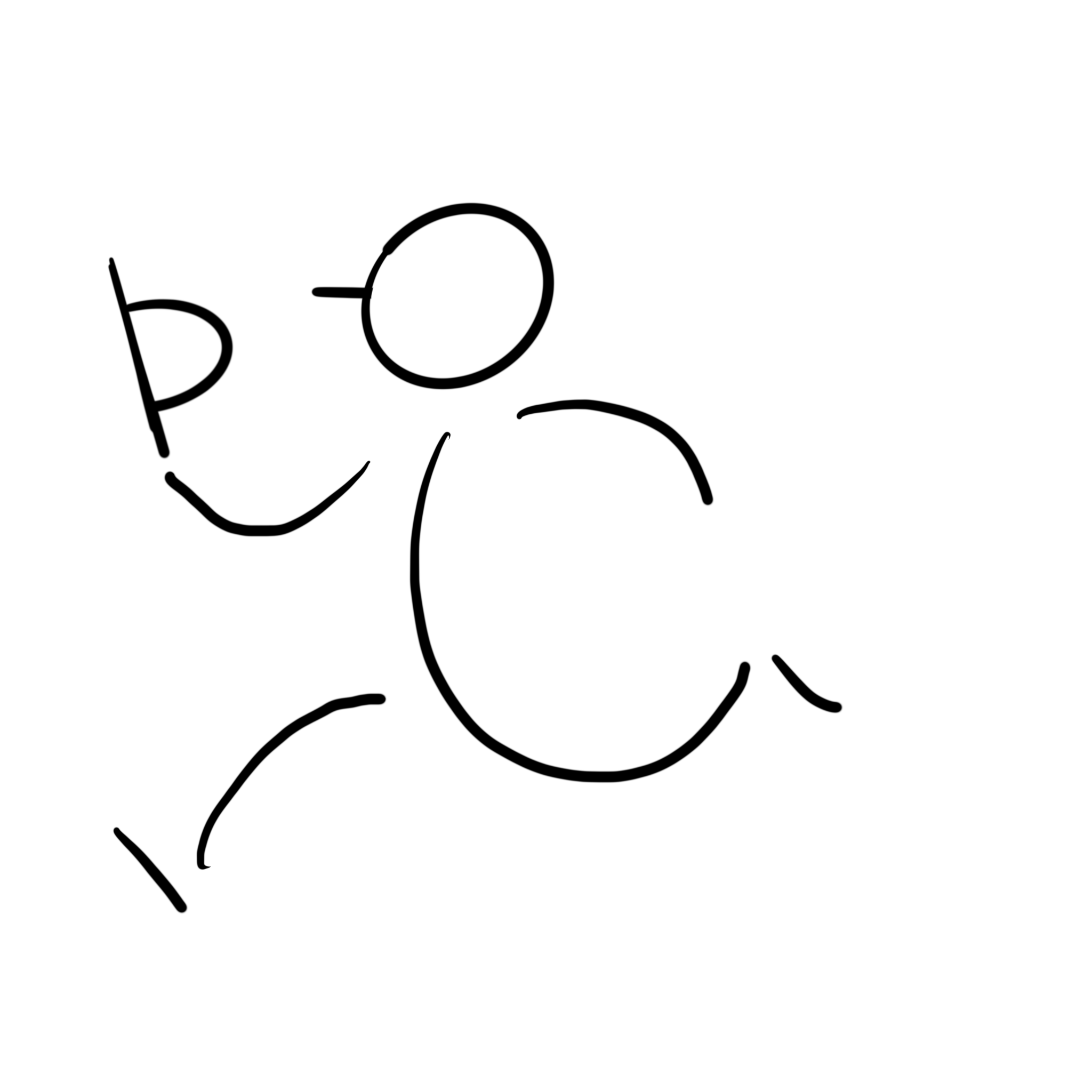}
\includegraphics[width=0.32\linewidth]{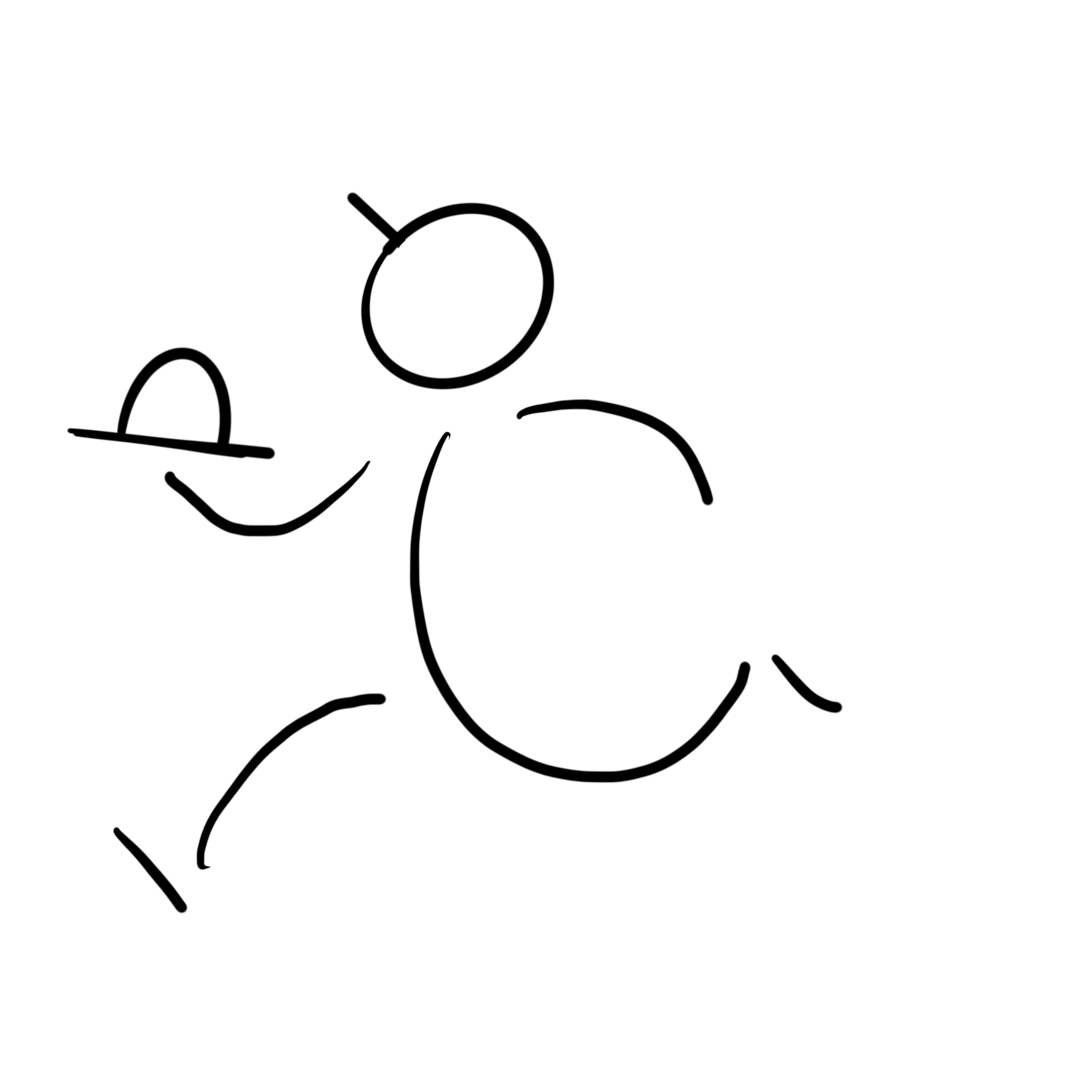}
\caption{Set-1: wire-person running holding a hat.}
\label{images/1}
\end{figure*}

\begin{figure*}[htb] 
\centering
\includegraphics[width=0.32\linewidth]{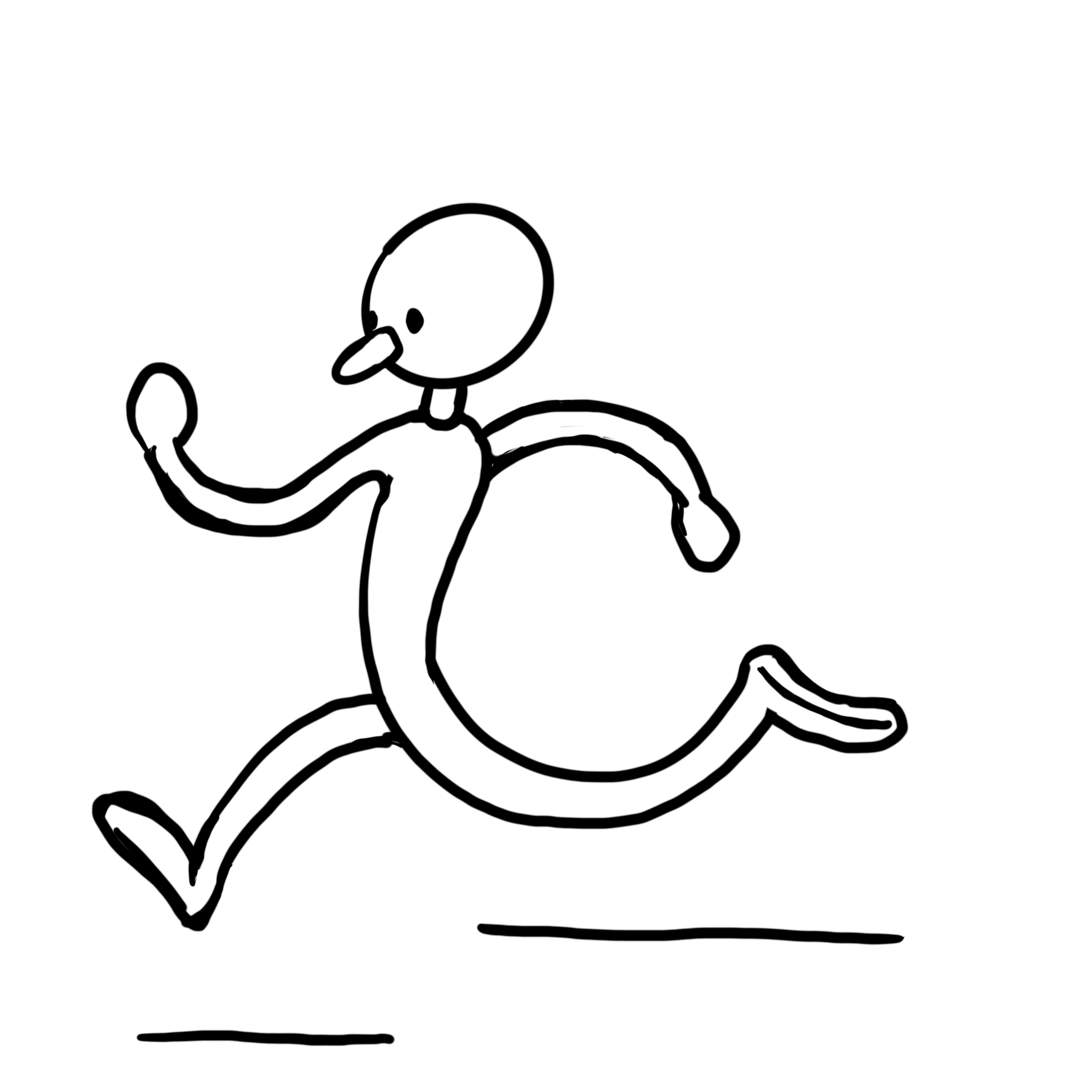}
\includegraphics[width=0.32\linewidth]{2/01}
\includegraphics[width=0.32\linewidth]{2/02}
\includegraphics[width=0.32\linewidth]{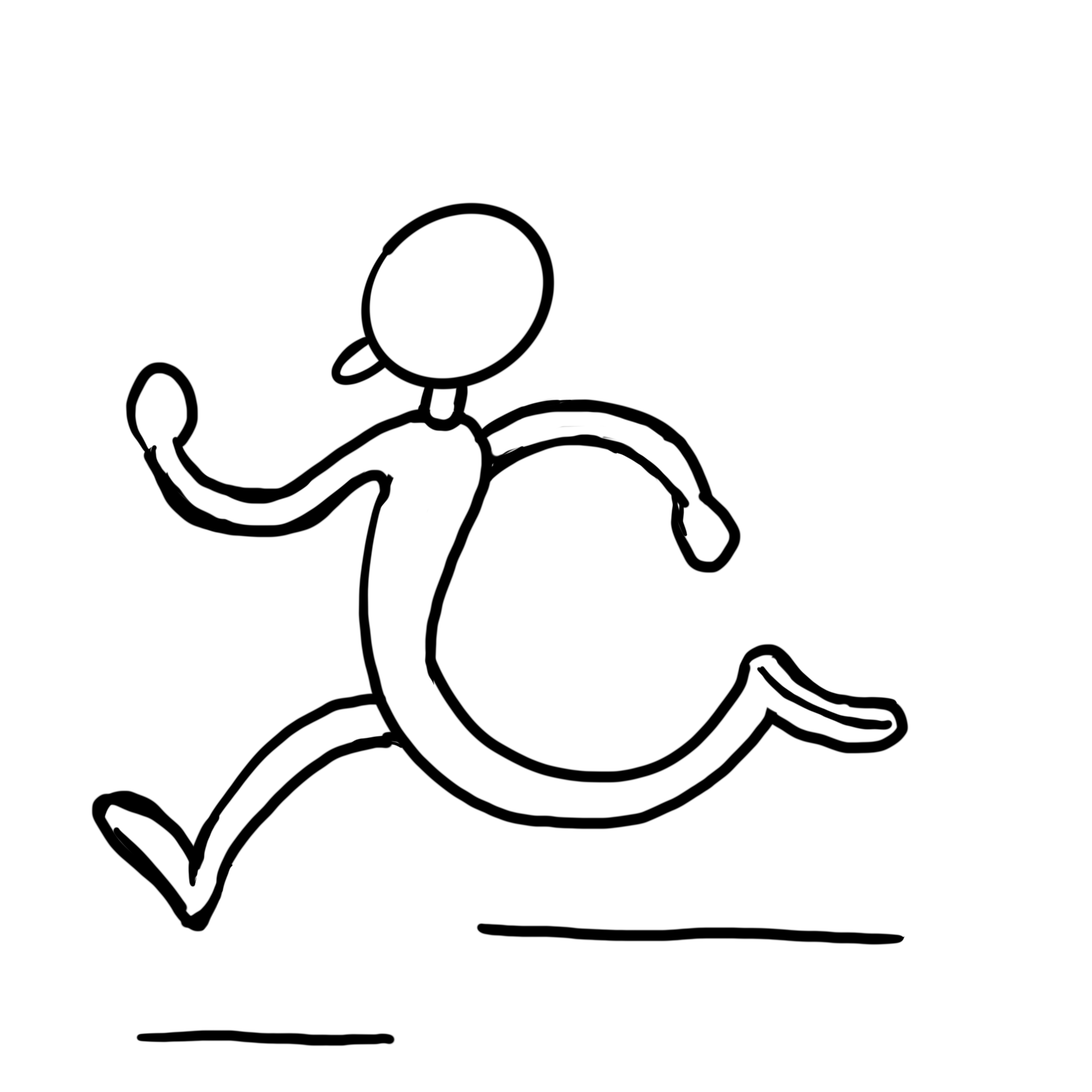}
\includegraphics[width=0.32\linewidth]{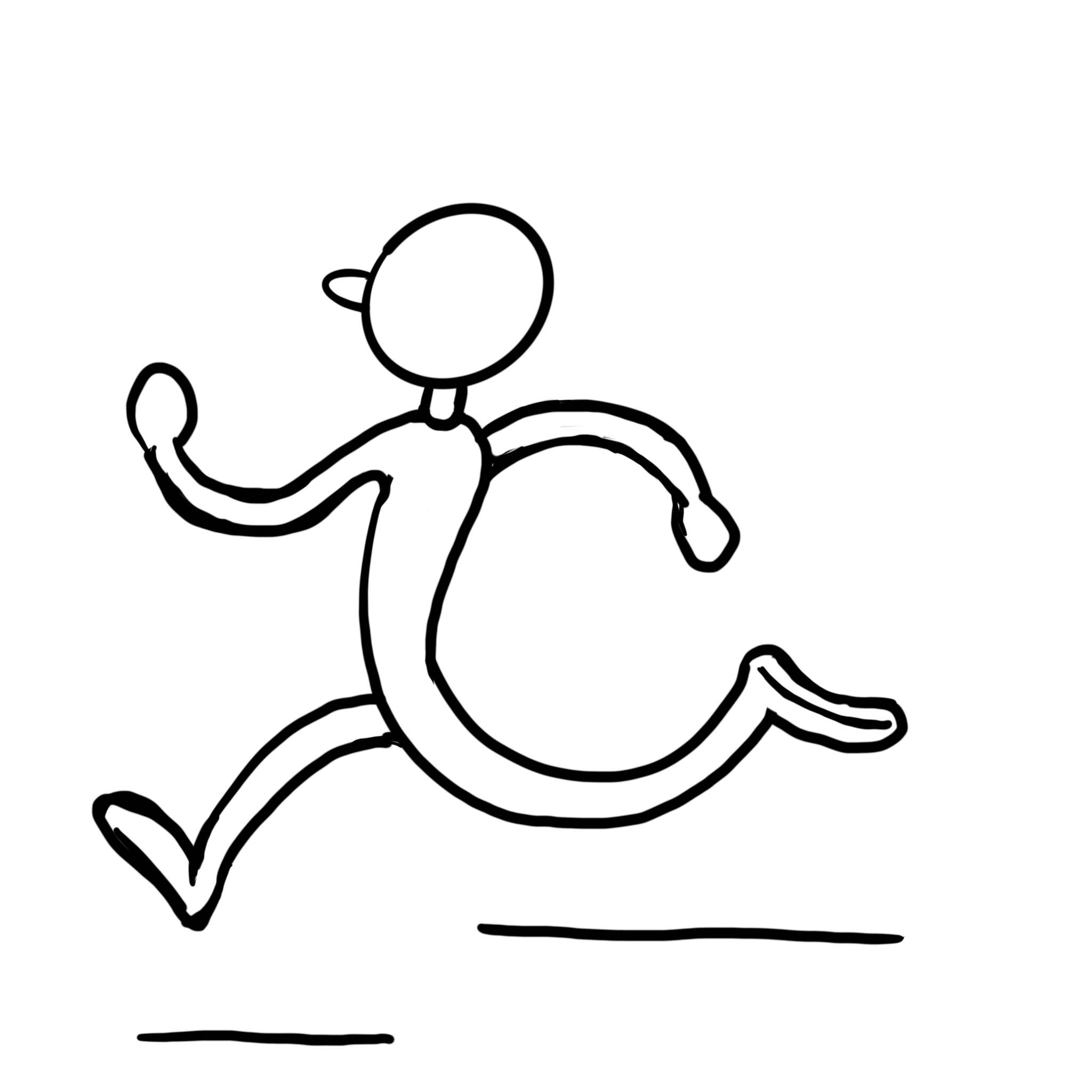}
\includegraphics[width=0.32\linewidth]{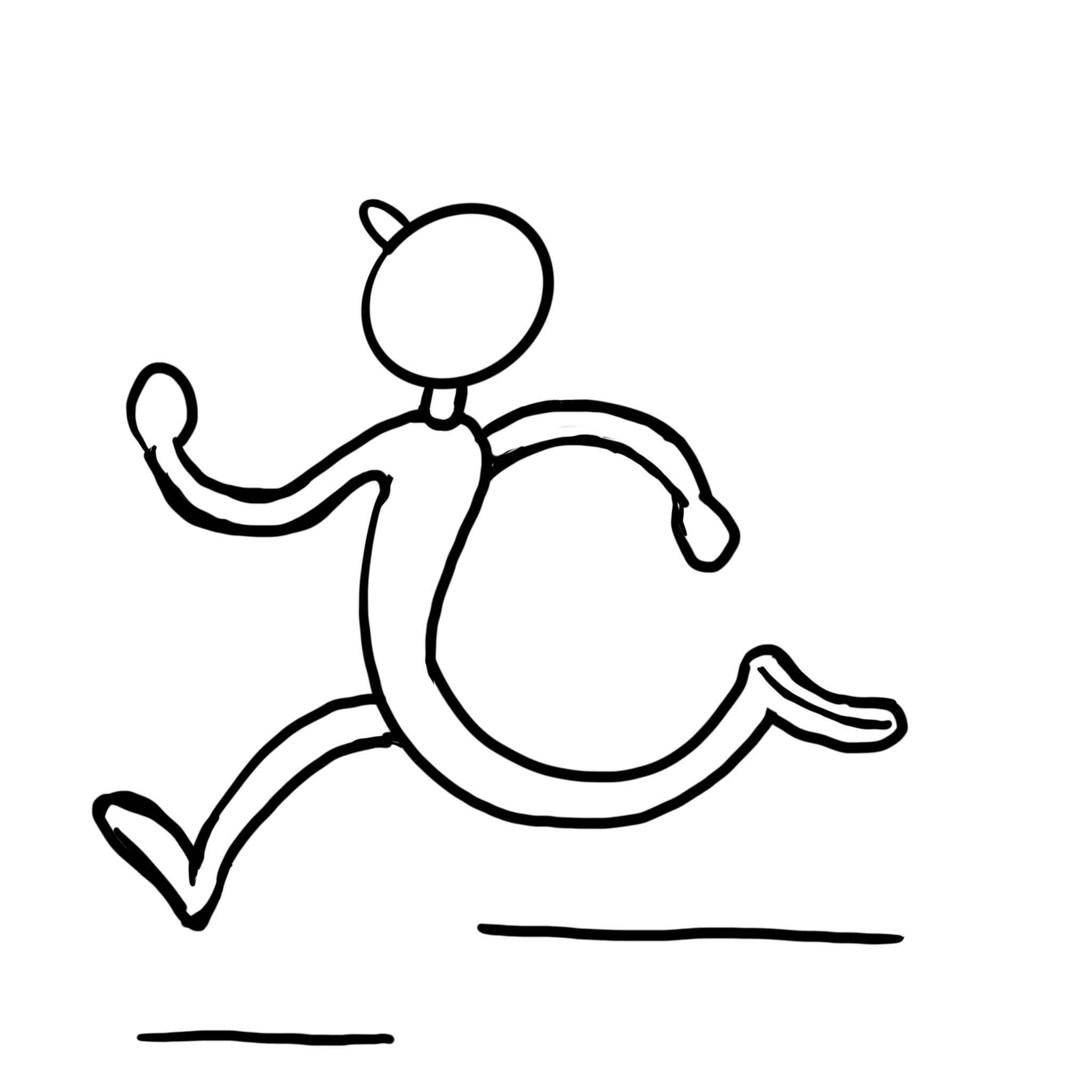}
\caption{Set-2: solid-person running, viewed from slightly behind viewpoint.}
\label{images/2}
\end{figure*}

\begin{figure*}[htb] 
\centering
\includegraphics[width=0.32\linewidth]{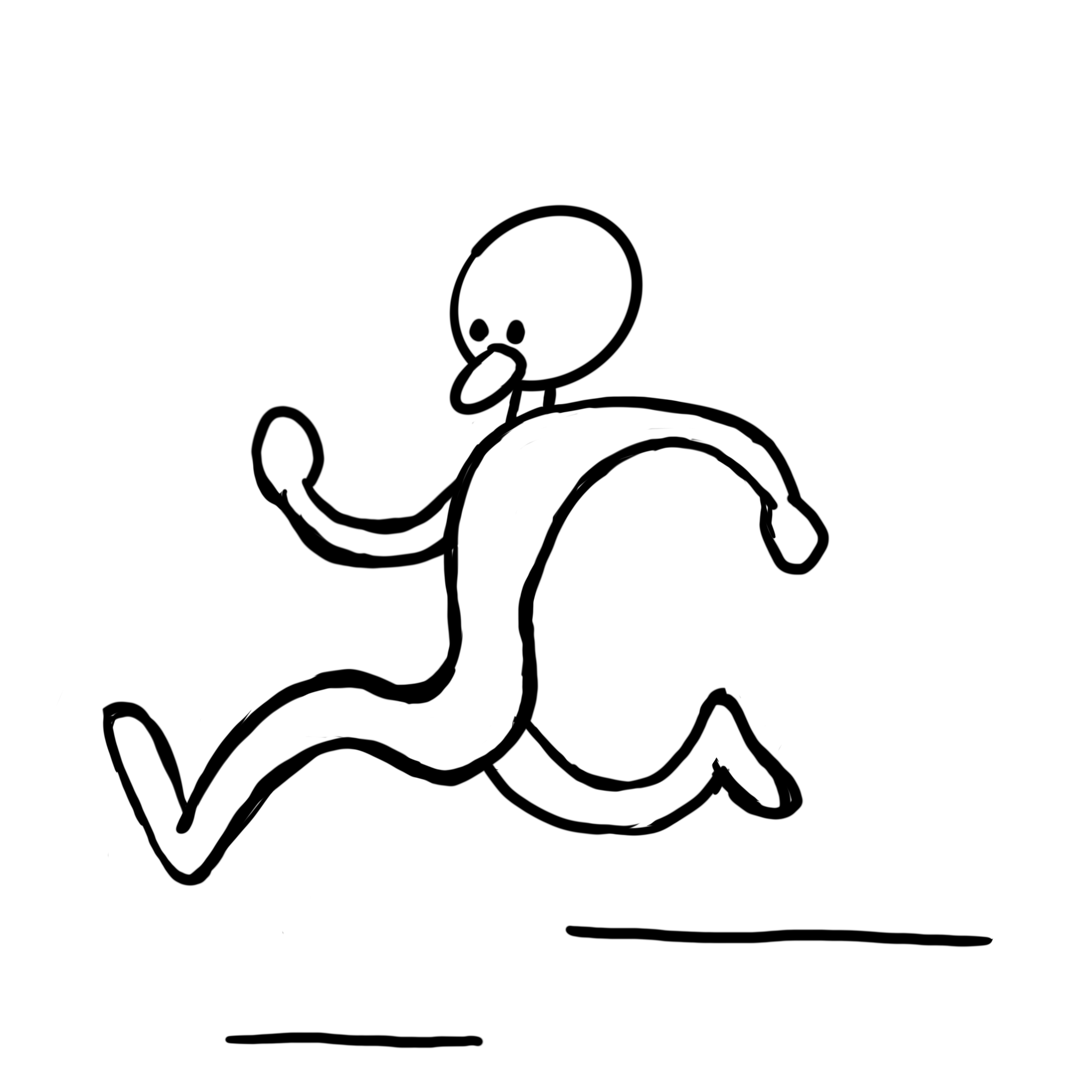}
\includegraphics[width=0.32\linewidth]{3/01}
\includegraphics[width=0.32\linewidth]{3/02}
\includegraphics[width=0.32\linewidth]{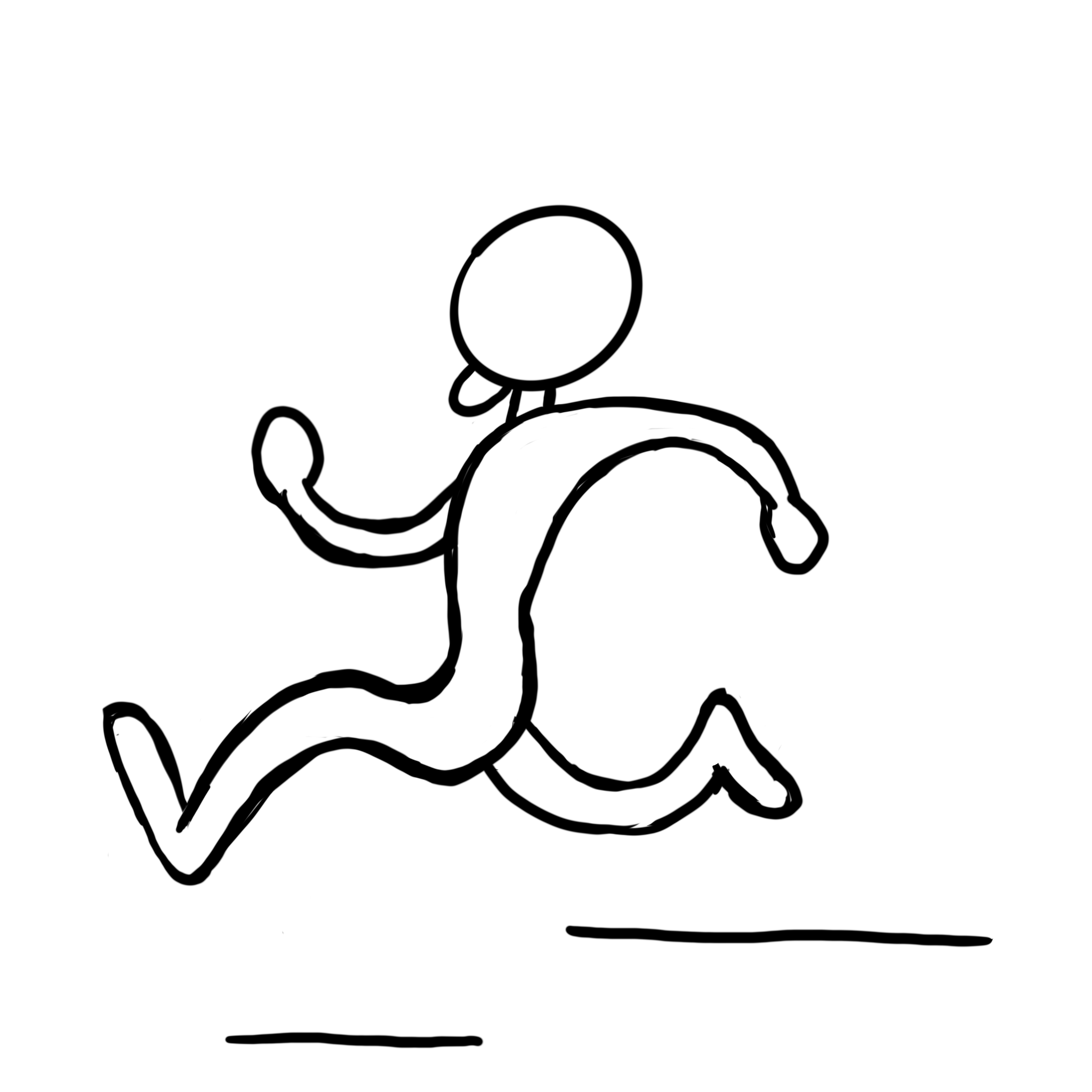}
\includegraphics[width=0.32\linewidth]{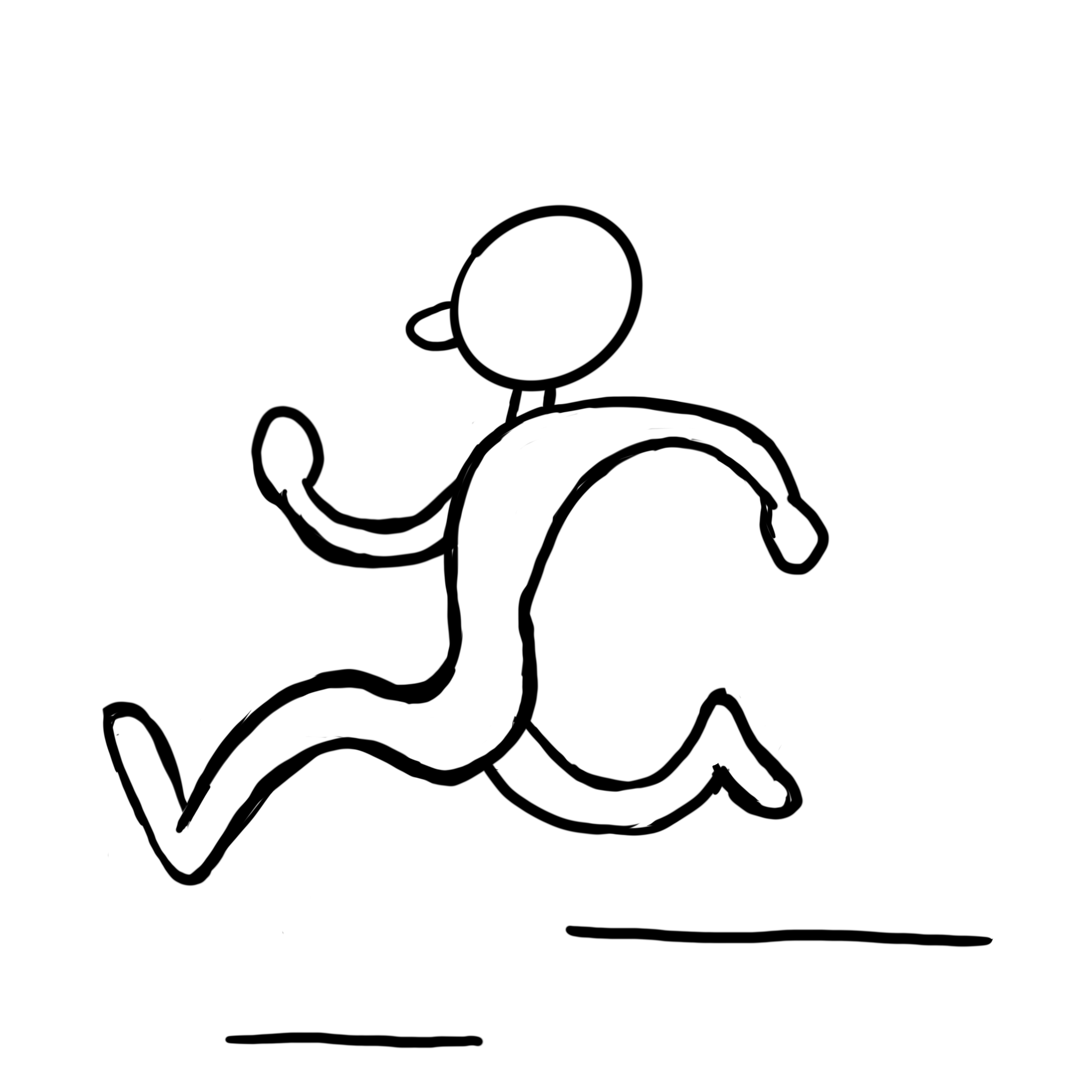}
\includegraphics[width=0.32\linewidth]{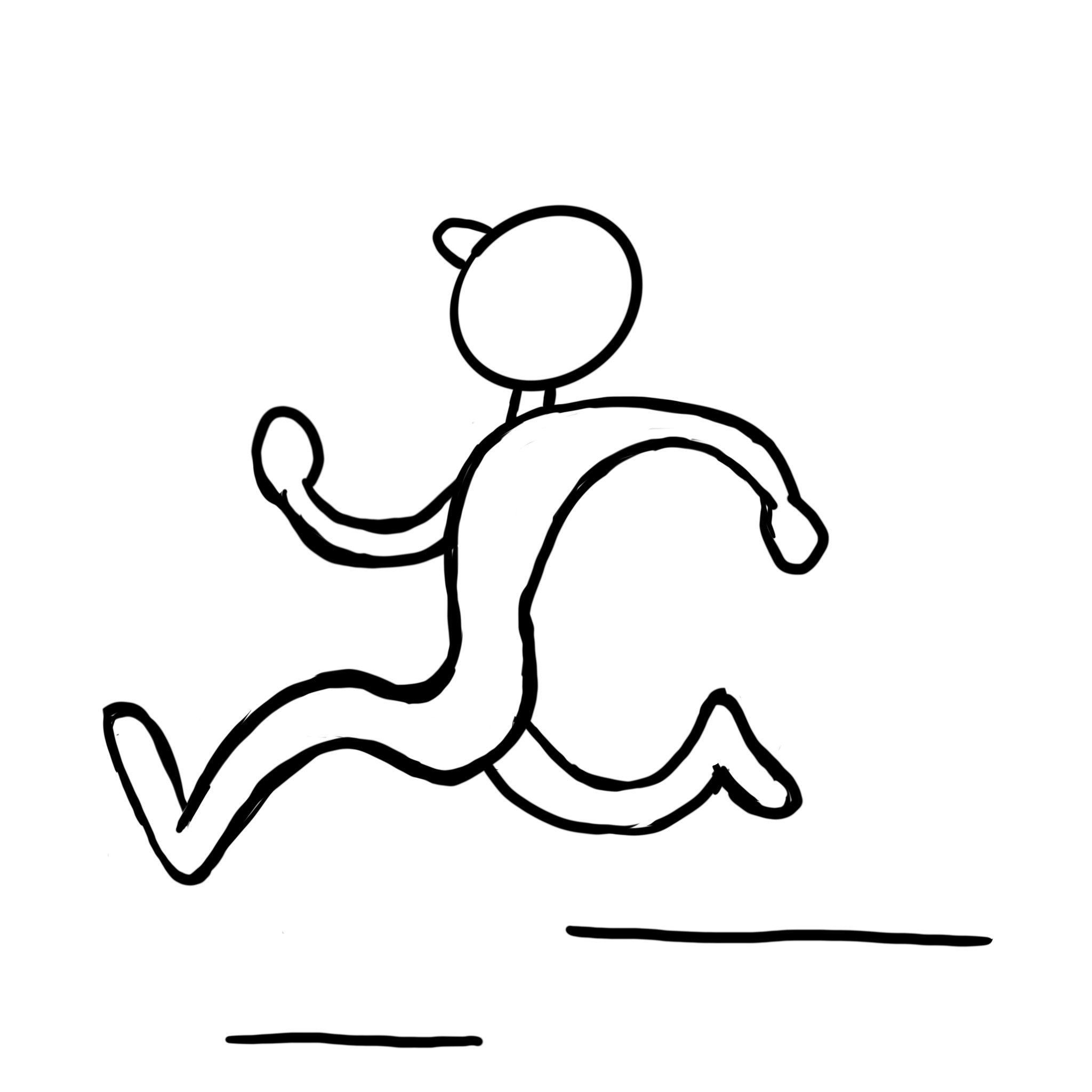}
\caption{Set-3: solid-person running, side view.}
\label{images/3}
\end{figure*}

\end{document}